\RequirePackage{fix-cm}
\documentclass[twocolumn]{svjour3}          
\smartqed  
\usepackage{graphics}      
\usepackage{graphicx}      
\usepackage{url}           
\usepackage{bm}            
\usepackage{subcaption}
\usepackage{float}
\usepackage{hyperref}
\usepackage{amssymb,amsmath}
\usepackage[numbers]{natbib}

\begin{document}

\title{Search for PeVatrons at the Galactic Center using a radio air-shower array at the South Pole
}


\author{A. Balagopal V.$^{*}$, A. Haungs, T. Huege, F. G. Schr{\"o}der 
}


\institute{
\email{aswathi.balagopal@kit.edu$^{*}$}\\
A. Balagopal V., F. G. Schr{\"o}der\at
Institut f{\"u}r Experimentelle Teilchenphysik, Karlsruhe Institute of Technology (KIT), Germany
 \and
 A. Haungs, T. Huege\at
 Institut f{\"u}r Kernphysik, Karlsruhe Institute of Technology(KIT), Germany
}

\date{Received: date / Accepted: date}

\maketitle

\begin{abstract}
The South Pole, which hosts the IceCube Neutrino Observatory, has a complete and around-the-clock exposure to the Galactic Center. Hence, it is an ideal location to search for gamma rays of PeV energy coming from the Galactic Center.
However, it is hard to detect air showers initiated by these gamma rays using cosmic-ray particle detectors due to the low elevation of the Galactic Center. 
The use of antennas to measure the radio footprint of these air showers will help in this case, and would allow for a 24/7 operation time.
So far, only air showers with energies well above $10^{16}$ eV have been detected with the radio technique.
Thus, the energy threshold has to be lowered for the detection of gamma-ray showers of PeV energy.
This can be achieved by optimizing the frequency band in order to obtain a higher level of signal-to-noise ratio. With such an approach, PeV gamma-ray showers with high inclination 
can be measured at the South Pole.
\end{abstract}

\section{Introduction}
\label{intro}
The study of air showers using radio detection techniques, to date, has been mainly applied in the case of charged cosmic-ray measurements and neutrino searches \cite{Huege:2016veh}\cite{Schroder:2016hrv}. 
Such showers have been detected with energy thresholds of at least a few tens of PeV. 
We show that this technique can also be used for PeV gamma ray astronomy,
by lowering the energy threshold. This can be done by extending the frequency band of measurement to higher frequencies than those used by current radio air shower arrays. 

The Galactic Center has been identified as a source of gamma rays of TeV energy by H.E.S.S. \cite{Abramowski:2016mir}. The source of this excess of TeV gamma rays has been traced to a PeVatron near the Galactic 
Center, 
in particular, close to the black hole Sgr A*. The H.E.S.S. data prefer a power law spectrum of E$^{-2.3}$ with no cut-off. The spectrum of TeV gamma rays can be extrapolated to PeV energies. 
Detection of PeV gamma rays approaching from the Galactic Center, would strengthen the evidence of the existence of such a PeVatron.

Current efforts to look for PeVatrons from the Galactic plane with the IceCube Observatory involve the measurement of the neutrino 
and muon fluxes from possible sources in the northern sky and the southern sky, respectively \cite{GonzalezGarcia:2009jc}\cite{Halzen:2009us}. 
Such searches then aim at explaining the contribution of PeVatrons to the knee of 
the cosmic-ray spectrum. 
Observing such PeVatrons using down-going muons will restrict the visible sky to that within the nearly vertical zenith angle range,
due to limitations in the detector volume. Hence, Galactic Center observations with the help of down-going muons would be restricted.

\begin{sloppypar}
Gamma rays of PeV energy, upon entering the Earth's atmosphere will produce air showers, similar to those produced by cosmic rays.
These air showers can be detected on the ground using particle detectors
and radio antennas. The Galactic Center is always visible at the South Pole, at an angle of $29^\circ$ above the horizon (zenith angle of $61^\circ$).
Hence the IceCube Observatory at the South Pole is an ideal location to search for gamma rays from the Galactic Center.
The number of gamma rays arriving at the IceCube Observatory from the Galactic Center with energies above 0.8 PeV is estimated to be around 11.5 events per year, 
from a simple extrapolation of the spectrum measured by H.E.S.S. (See appendix~\ref{sec:flux}).
\end{sloppypar}

The IceCube Neutrino Observatory \cite{Achterberg:2006md}, the 1 km$^3$ array for the detection of astrophysical neutrinos, has a surface component of ice-Cherenkov particle detectors (IceTop) used for the detection of cosmic-rays \cite{IceCube:2012nn}.
It is planned to upgrade IceTop using scintillators \cite{HuberSam}. It is also foreseen to have a large surface array of scintillation detectors
and air-Cherenkov telescopes
as a part of IceCube-Gen2 \cite{JvanSanten}
\cite{2015ICRC...34.1070E}\cite{Seckel:2015aex}\cite{IceAct2017}.
A surface array of radio antennas in addition
to this could potentially increase the accuracy for the detection of air showers (especially inclined air showers) and for the determination of mass composition. 
RASTA, 
a previous study that was made using test antennas
at the South Pole explored the possibility of improving veto capabilities and cosmic-ray studies at IceCube \cite{Boser:2010sw}. 
Apart from this, in-ice radio measurements at the South Pole is also being made using ARA, that aims at
measuring the radio signals produced by high energy neutrinos inside ice \cite{Allison:2015eky}.


%
\begin{figure*}[tbp]
  \begin{subfigure}[b]{.49\linewidth}
  \includegraphics[width=8cm]{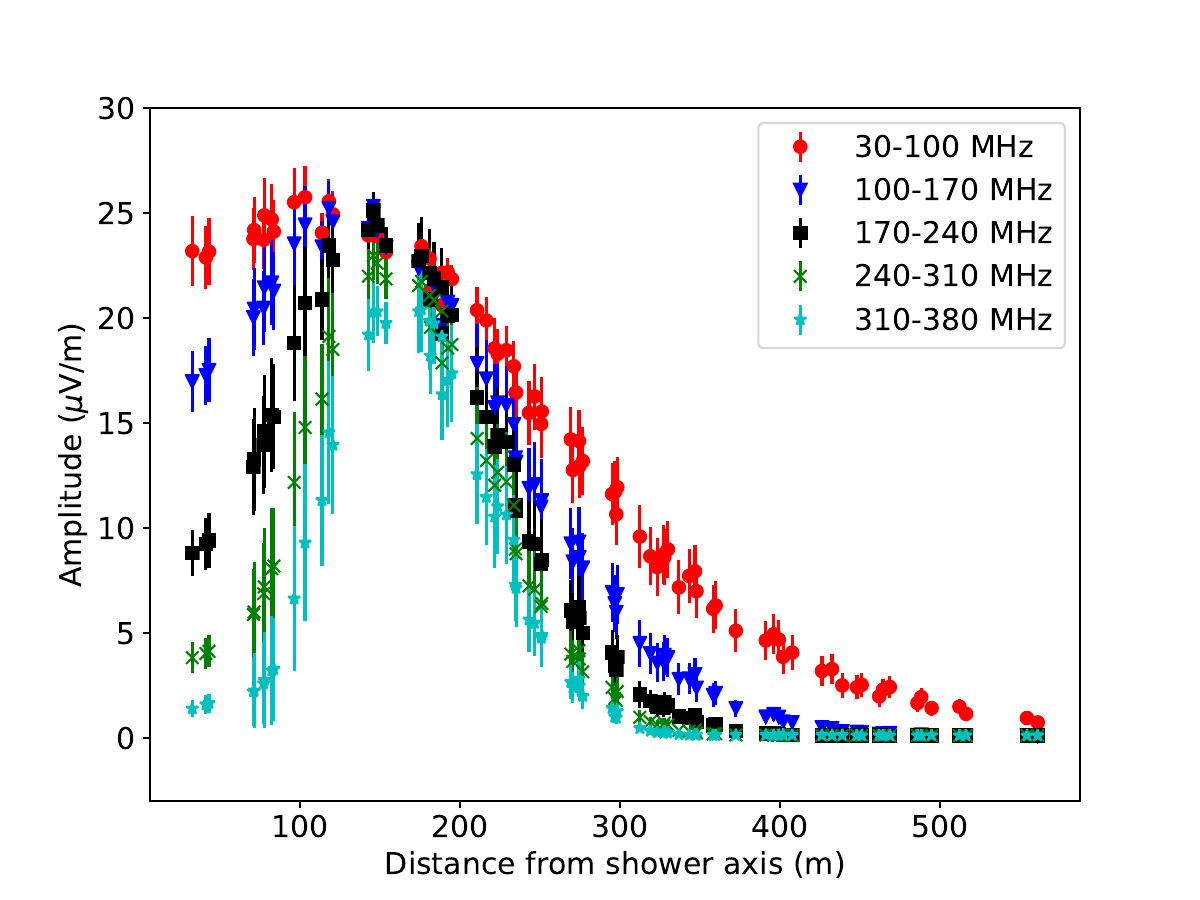}
 \caption{10 PeV gamma-ray showers, $\theta = 61^{\circ}, \alpha =  79^{\circ}$} 
  \end{subfigure}
  \begin{subfigure}[b]{.49\linewidth}
  \includegraphics[width=8cm]{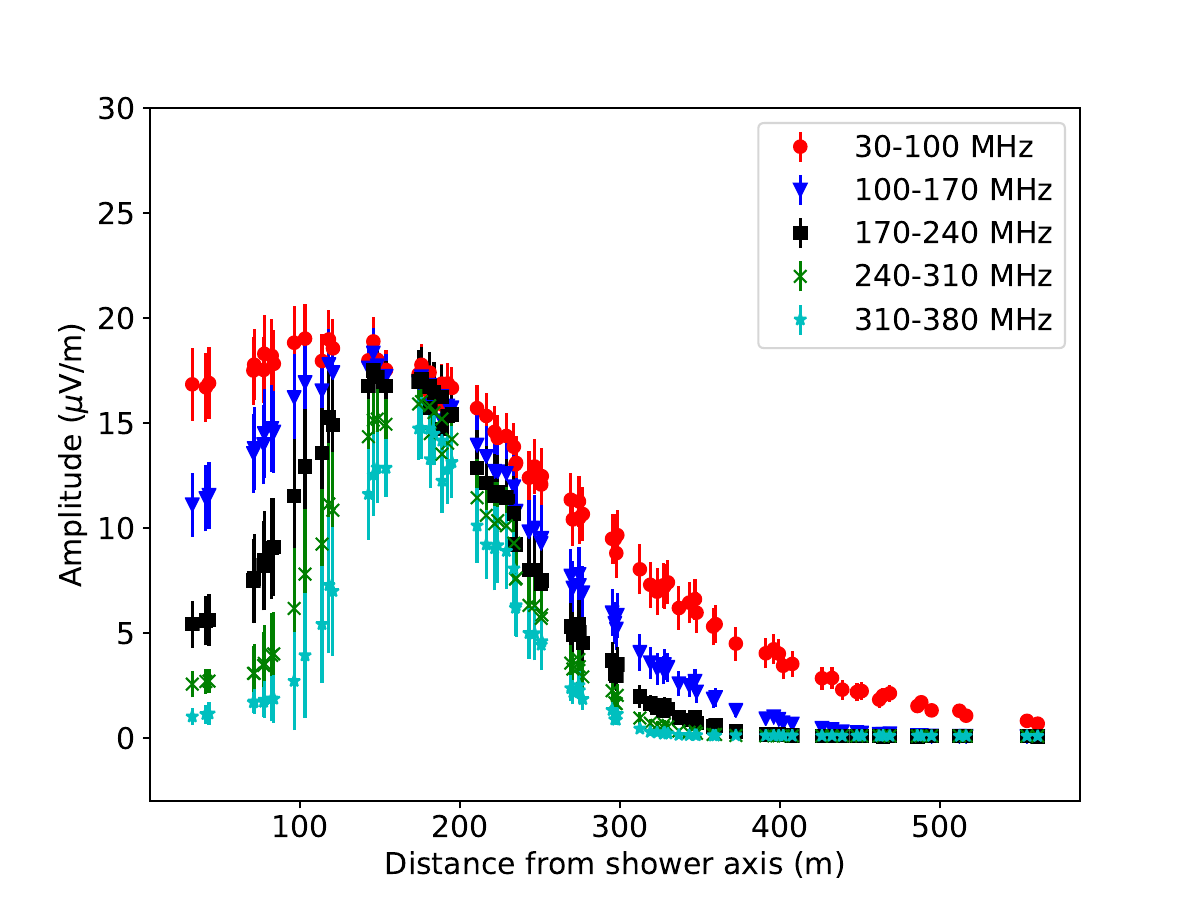}
 \caption{10 PeV proton showers, $\theta = 61^{\circ}, \alpha =  79^{\circ}$}

  \end{subfigure}
 \caption{Lateral distribution of radio signals for gamma-ray and proton showers for frequencies from 30 MHz to 380 MHz. For illustration the signals are shown in various frequency bands with a width of 70 MHz each. \label{fig:ldf}}

\end{figure*}





A surface array of radio antennas at the South Pole can be used to search for air showers produced by PeV gamma rays arriving from the Galactic Center. 
Inclined air showers of PeV energy will be hard to detect and reconstruct effectively using particle detectors, since a major part of the shower dies out by the
time it reaches the detector array. This is especially the case for showers induced by gamma rays. Gamma-ray showers have lesser muonic content when compared to hadronic showers. 
In particular, the showers induced by protons have a significantly larger fraction of muons than gamma-ray showers. 
Muons, unlike electrons and 
positrons,
are the most prominent component of inclined air showers of PeV energy, that will reach the ground. The low muonic content of gamma-ray showers results in fewer pulses in the IceTop tanks and the future scintillation detectors. 
In contrast, the radio signal from a shower with the same primary energy survives and can be detected on the ground
by an array of radio antennas. Thus by comparing the radio emission to the number of muons detected on the ground one can distinguish between showers initiated by gamma rays and those by other nuclei.

\begin{sloppypar}
Radio emission of air showers develops mainly due to the deflection of the electrons and positrons of the shower in the Earth's magnetic field (Geomagnetic effect). This results in a time-varying current that 
produces radio pulses \cite{Kahn206}\cite{Scholten:2007ky}. 
Another contribution to radio emission comes from the Askaryan effect, which is due to the charge excess at the shower front that forms as the shower propagates through the atmosphere 
\cite{Askaryan_1961}\cite{Askaryan_1965}. This effect, which has a 
smaller influence
in air showers than the geomagnetic effect, causes a small asymmetry in the total radio emission. At higher frequencies, a Cherenkov ring is visible in the radio footprint, due to the time compression of radio pulses 
caused
by the refractive index of air \cite{deVries:2011pa}. 
For example, at frequencies such as that covered by the band
50-350 MHz, which is used by 
 the lower frequency component of the Square Kilometer Array (SKA-LOW) \cite{7928622}, the Cherenkov ring is visible.
Such a Cherenkov ring is only marginal in the frequency band  
30-80 MHz, which is the frequency range used by most of the existing radio air shower experiments,
e.g. AERA \cite{1748-0221-7-10-P10011}, Tunka-Rex \cite{Bezyazeekov:2015rpa} and LOFAR \cite{Schellart:2013bba}.
\end{sloppypar}

An inclined shower produced by a gamma ray from the Galactic Center will leave a large radio footprint on the ground, whose diameter ranges from several 100 m to km depending on the angle of inclination.
Recent studies of inclined air showers by The Auger Engineering Radio Array (AERA) have 
experimentally proven
this \cite{Kambeitz:2016rqu}. The footprint detected on the ground is elliptical in
shape because of pure geometrical reasons \cite{Huege:2015lga}. The inclined air showers detected by AERA have energies higher than $10^{18}$ eV. Similar characteristics will be seen by showers of PeV
energy arriving at the IceCube location. 

Figure~\ref{fig:ldf} shows the different simulated amplitudes delivered to 81 antennas, on an area of 1 km$^2$, at the location of the IceTop stations \cite{IceCube:2012nn}; that is with one antenna placed 
at the center of the two Cherenkov tanks that form an IceTop station (see also Figure~\ref{fig:SNRmap}).
The lateral distribution of the amplitudes in the figure are those from gamma-ray induced showers and proton induced showers at these antenna locations. 
Here, the zenith angle is fixed to $61^\circ$. These showers have a geomagnetic angle (angle between the 
Earth's magnetic field and the shower axis) of $\alpha = 79^\circ$.
For illustration, the frequencies are split into bands with a width of 70 MHz each, and range from 30 MHz all 
the way up to 380 MHz. 
The proton showers have lower amplitudes than gamma-ray showers since they have lower electromagnetic content. The plot shows the mean amplitude along with the spread about the mean
with 30 simulated showers for gamma-ray and proton primaries. 
We can see that the the lateral distribution for these showers change with the frequency of observation. At frequencies above 100 MHz, we start to 
see the Cherenkov ring in such a distribution. At very high frequencies like those above 300 MHz, the emission becomes extremely localized, giving
non-zero values of the amplitude only on the Cherenkov ring.
That is, the radio signal dies out gradually at every other location at such high frequencies.

So far, it has been considered that air showers from cosmic rays with an energy range greater than $10^{16}$ eV can be measured using the radio detection technique. At energies lower than this, the background 
overwhelms
the radio signal from an air shower \cite{Huege:2016veh}. This is especially the case for the frequency range of 30-80 MHz. This makes it hard to measure air showers at low energies, unless interferometric methods are used. Thus in order to
measure  air showers produced by PeV gamma rays, we explore a different method to lower the energy threshold. This paper focuses on this aspect, especially on the optimization of the observing frequency bands in order to lower the threshold
energy for the detection of gamma-ray air showers.
\section{Simulation of air showers}
\begin{sloppypar}
A thorough study of the air showers that are produced by the incoming gamma rays is needed for predicting the radio signal that will be detected on the ground. For this purpose, air shower simulations were performed
using CoREAS \cite{Huege:2013vt}, which is the radio extension of CORSIKA \cite{Heck:1998vt}. We use CORSIKA-7.4005 with hadronic interaction models FLUKA-2011.2c.2 and SIBYLL-2.1 \cite{PhysRevD.80.094003}.
Later simulations used CORSIKA-7.5700 with SIBYLL-2.3. This was not seen to change the received radio signals from air showers significantly (since the main difference between the versions is in the muonic content
of the hadronic interactions).
A total of 1579 simulations have been done for this study.
The simulations used the atmosphere of the South Pole (South pole atmosphere for Oct. 01, 1997 provided in CORSIKA) with an observation level of 2838 m above the sea level. 
All the showers have been simulated using the thinning option (with a thinning level of $2.7 \times 10^{-7}$).
\end{sloppypar}

The showers simulated are those of gamma-ray primaries with energies ranging from 1-10 PeV. The azimuth angle for preliminary studies were fixed so that the shower axis is oriented anti-parallel
to the Magnetic North ($\phi$ = 0), thereby giving 
a geomagnetic angle of $79^\circ$ for a shower with a zenith angle of $61^\circ$.
At the South Pole, the magnetic field is inclined at an angle of $18^\circ$
with respect to the vertical, with an intensity of 55.2 $\mu$T. The zenith angle is
fixed to $61^\circ$ for a major portion of the simulations since this is the inclination of the Galactic Center at the South Pole. The core position was set at the center of the IceTop array, i.e. at (0,0).
For comparison, proton showers were also simulated, with the same parameters. The simulations included 81 antennas, each at the center of an IceTop station. This resulted in an array where the average antenna
spacing is around 125 m. An inner infill array of antennas with much denser spacing (approximately 90 m) is also present, since such a structure is present for the IceTop stations also. The entire array covers 
an area of
around 1 km$^2$.

The output from CoREAS simulations gives the signal strength at each of these antenna stations in units of $\mu$V/m. This has to be folded through the response of an antenna, in order to estimate the measurable 
signal.
For this purpose, a simple half wave dipole antenna with resonance at 150 MHz was simulated using NEC2++ \cite{necpp}. Antennas in the east-west and in the north-south direction (with respect to the magnetic field)
were used
at the location of each station, to extract the complete signal from the air shower. Here, the z-component is neglected. This can be safely done because of the small angle between the magnetic field and the 
vertical, thereby resulting in a smaller z component of electric field as compared to the x and y components, even for the inclined air showers.

The simulations that are performed here are simplified, since the main focus is to understand the required experimental setup for lowering the energy threshold in order to detect PeV gamma rays from the Galactic Center.
Specific effects like the impact of an optimized type of antenna or other details that can be important for a particular experimental setup are ignored in this context and can be included in the case of a  more detailed study.

\begin{figure}[tbp]
\centering
\hspace{-0.3cm}
\begin{subfigure}[b]{.99\columnwidth}
  \includegraphics[height=.69\columnwidth]{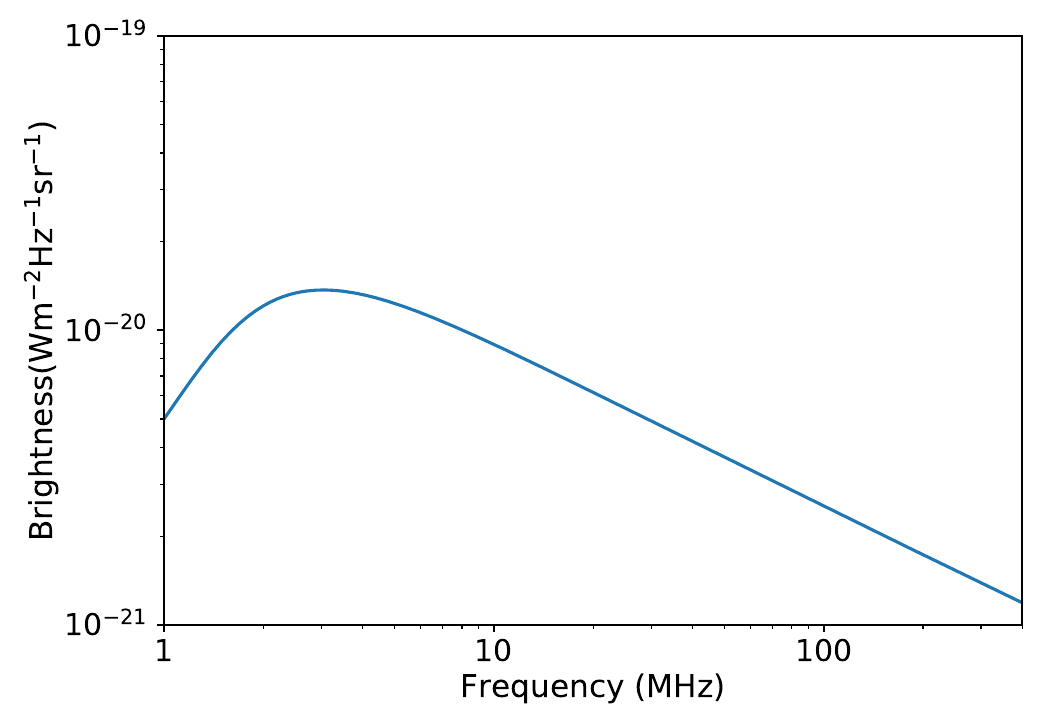}
\label{fig:Cane(a)}
\vspace{-0.3cm}
\caption{}
\end{subfigure}

\centering
\begin{subfigure}[b]{.99\columnwidth}
  \includegraphics[height=.75\columnwidth]{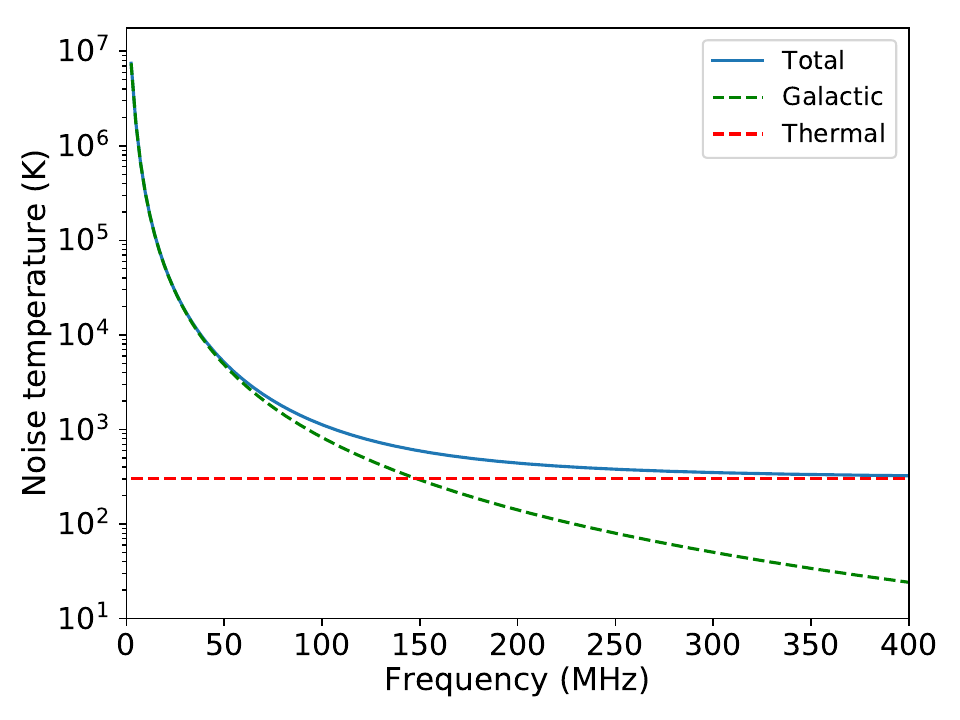}
\label{fig:Cane(b)}
\vspace{-0.3cm}
\caption{}
\end{subfigure}
\caption{(a) Brightness of the Galactic radio background radiation  given by the Cane function \cite{Cane}. (b) Total noise temperature as a function of frequency. \label{fig:Cane}}

\end{figure}
\begin{figure*}[tbp]
  \begin{subfigure}[b]{.49\linewidth}
 \includegraphics[width=8cm]{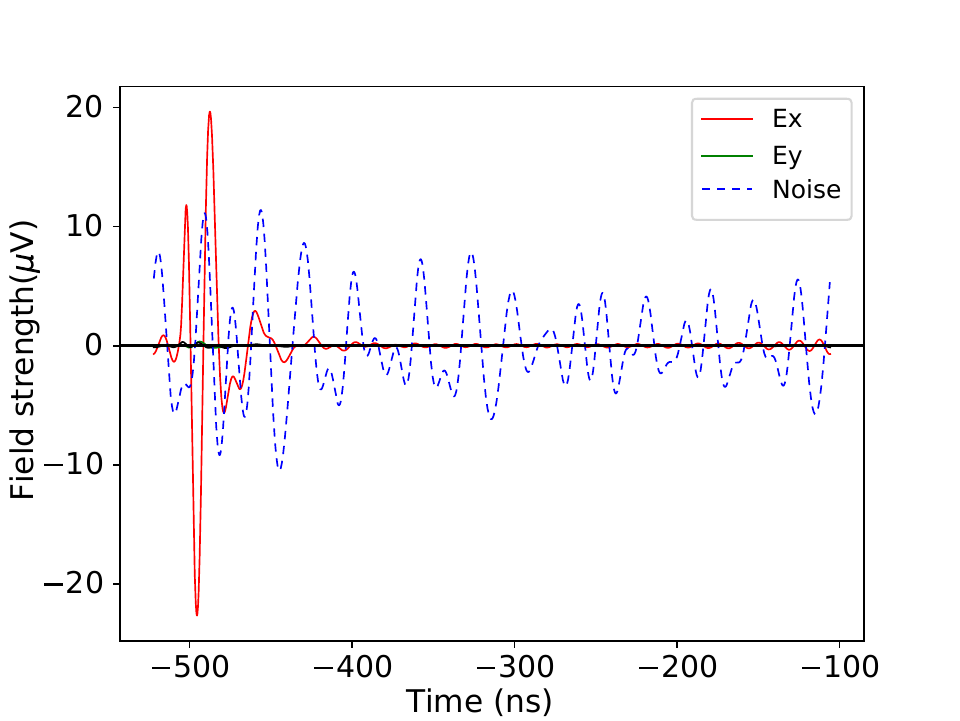}
 \caption{30-80 MHz, (SNR $\approx$ 35)} 
  \end{subfigure}
  \begin{subfigure}[b]{.49\linewidth}
   \includegraphics[width=8cm]{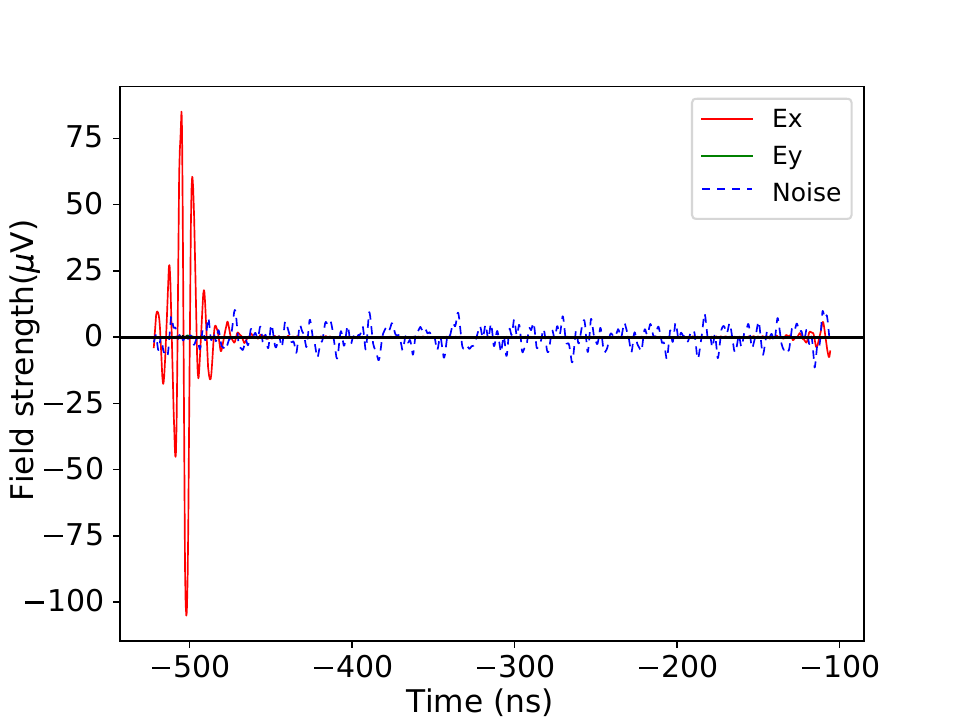}
 \caption{50-350 MHz, (SNR $\approx$ 1055)} 

  \end{subfigure}
 \caption{Simulated signal for a 10 PeV $\gamma$-ray shower and noise at a station located on the Cherenkov ring, with a distance of 107 m from the shower axis. 
 The y-component of the signal shown here is nearly zero. 
 The received noise is much lower at higher frequencies, leading to a higher signal-to-noise ratio.  }
  \label{fig:traces}

\end{figure*}
\section{Inclusion of a noise model}

One of the major challenges for the detection of radio signals from showers of PeV energy is the lower level of signal, when compared to the background noise as discussed in section \ref{intro}. There can be external as well as internal (thermal) 
sources of noise
for radio air shower experiments. 
The external sources of noise range from Galactic noise through man-made noise to noise contributed by atmospheric events. At the South Pole, the external contribution mainly comes from Galactic noise as the 
contributions from  other elements are expected to be much lower in comparison.
Existing air shower experiments point out that measurement of signals from inclined air showers of PeV energy range using the band of 30-80 MHz is hard to achieve. Within this frequency band, the signal will be 
dominated by noise, especially for showers in the PeV energy range. 

In this study, a simplified and average model of diffuse Galactic noise developed by Cane \cite{Cane} is used. 
It has already been shown by measurements from RASTA and ARA that the Cane model describes the Galactic noise measured at the South Pole with a reasonable accuracy \cite{MikeRATSA}\cite{Allison:2011wk}.
The noise is given in units of brightness (Galactic Brightness background) in this model, as can be seen in Figure~\ref{fig:Cane}. 
The corresponding 
brightness temperature, obtained from the relation $T = \frac{1}{2k_{\mathrm{B}}}\frac{c^{2}}{\nu^{2}} B(\nu)$, is used for determining the Galactic contribution to the total noise. In addition to the Galactic noise, there is also 
a contribution from the thermal component, which arises due to the electronic boards
and other equipments related to the experimental setup (internal noise). A thermal noise of 300 K is used here. 
With a very simple hardware, the thermal noise contribution could even be more than 300 K.
Much lower noise levels of a few 10 K can be achieved by dedicated hardware optimization. 

The noise temperature can be related to the power received in the antenna by $P = k_{\mathrm{B}} T \delta\nu$, where $\delta\nu$ is the frequency interval within which the power is extracted
and $k_{\mathrm{B}}$ is the Boltzmann constant. From  Figure~\ref{fig:Cane},  we see that the Galactic 
noise diminishes as the 
frequency increases. At frequencies above $\approx$ 150 MHz, we become mainly limited by the thermal noise.

The expected noise for a given frequency band can be expressed as time traces 
from the predicted noise temperature within this band (See appendix~\ref{sec:noisetrace}). Noise traces extracted like this can be compared to signals from air showers as shown in Figure~\ref{fig:traces}. 
Here, the signals considered are those 
for a gamma-ray shower of 10 PeV energy and inclined at an angle of $61^\circ$.

It is clearly seen that the signal-to-noise ratio increases as we move to higher frequencies, as is expected from the behavior of the Galactic noise. 
Here, the signal-to-noise-ratio is determined as  $\mathrm{SNR} =  S^2/N^2$
where $S$ is the maximum of the Hilbert envelope over the signal and $N$ is the RMS noise in the specified frequency band.  
The suppression of the Galactic noise beyond 150 MHz is visible in the time traces of the noise. It becomes obvious that moving on to higher frequencies will enable us to have a higher level of signal-to-noise 
ratio (SNR), provided the antenna falls within the footprint of the shower.


\section{Optimizing the observing frequency band}
Although it is clear from Figure~\ref{fig:traces} that using frequency bands that are higher than the standard band (30-80 MHz) will help us in enhancing the signal-to-noise ratio, the exact band 
that should be used for maximizing the chances of observation is still unclear.
It is of course, possible to measure in wide band
frequencies, and then to digitally filter into the required frequency range. But this will increase the cost of the experiment considerably, since the usage of higher frequencies require a greater sampling rate and 
hence better
communication facility, memory, ADC, etc. Thus, a detailed study is made to estimate the frequency range that will give a maximum signal-to-noise ratio (and thereby maximize the detection probability),
and can hence be used for the experiment, which is the 
focus in the following section.

A close inspection of the shower footprint at higher frequencies 
reveals that there are three regions of interest: on the Cherenkov ring, inside the Cherenkov ring, and 
outside the 
Cherenkov ring. It is desirable to have a high value of SNR in all of these regions for maximizing the probability of detection
in the entire antenna array. 


\begin{figure}
\centering
 \captionsetup[subfigure]{justification=centering}

 \begin{subfigure}{.99\columnwidth}

\includegraphics[width=8.1cm]{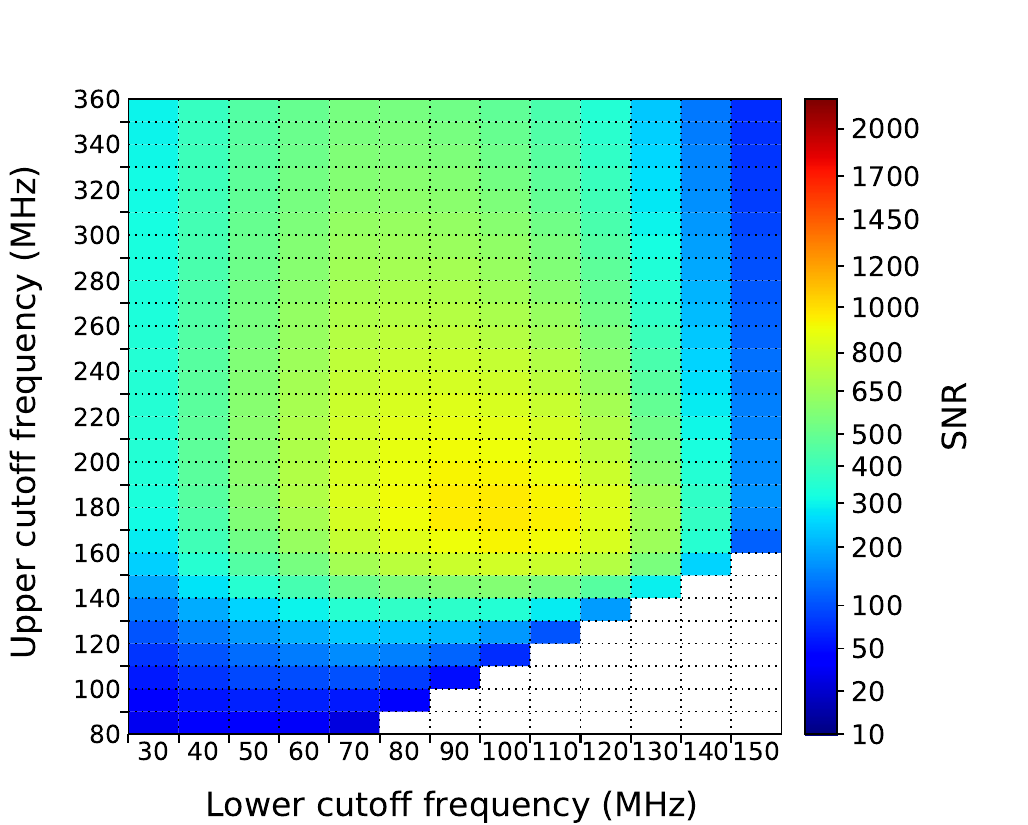}
\caption{Inside the Cherenkov ring (distance: 34 m)}
\end{subfigure}
 \begin{subfigure}{.99\columnwidth}
  \includegraphics[width=8.1cm]{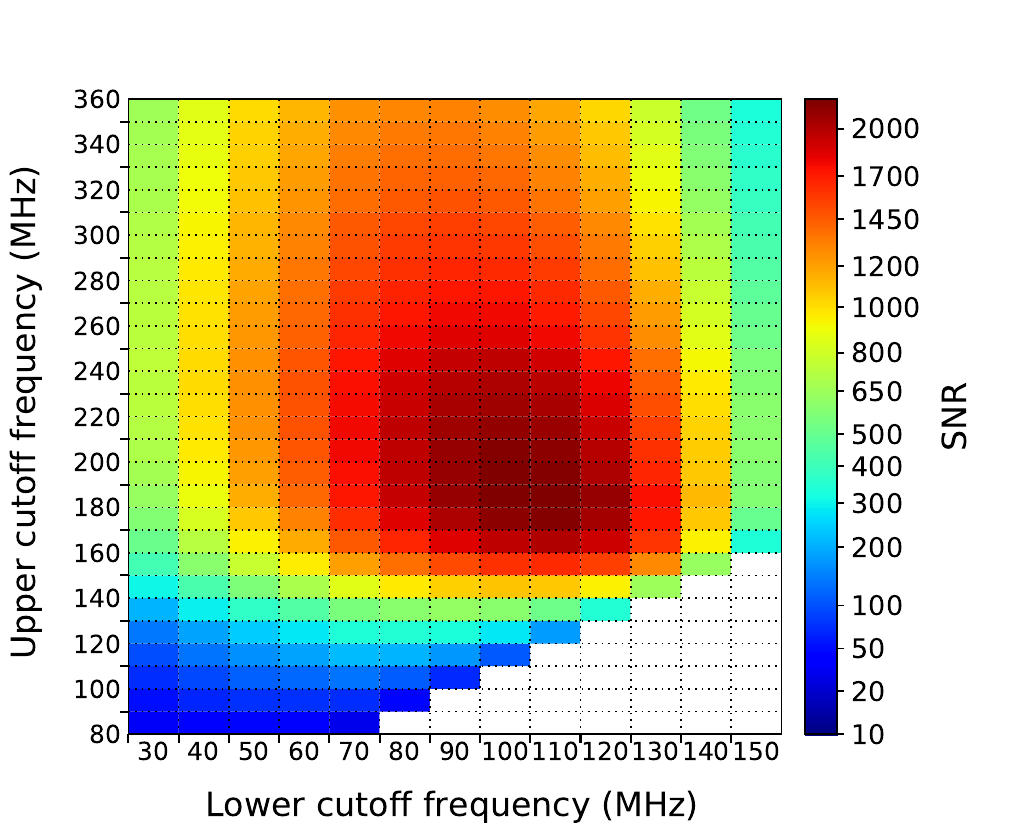}
 \caption{On the Cherenkov ring (distance: 107 m)}
 \end{subfigure}

\begin{subfigure}{.99\columnwidth}
 \includegraphics[width=8.1cm]{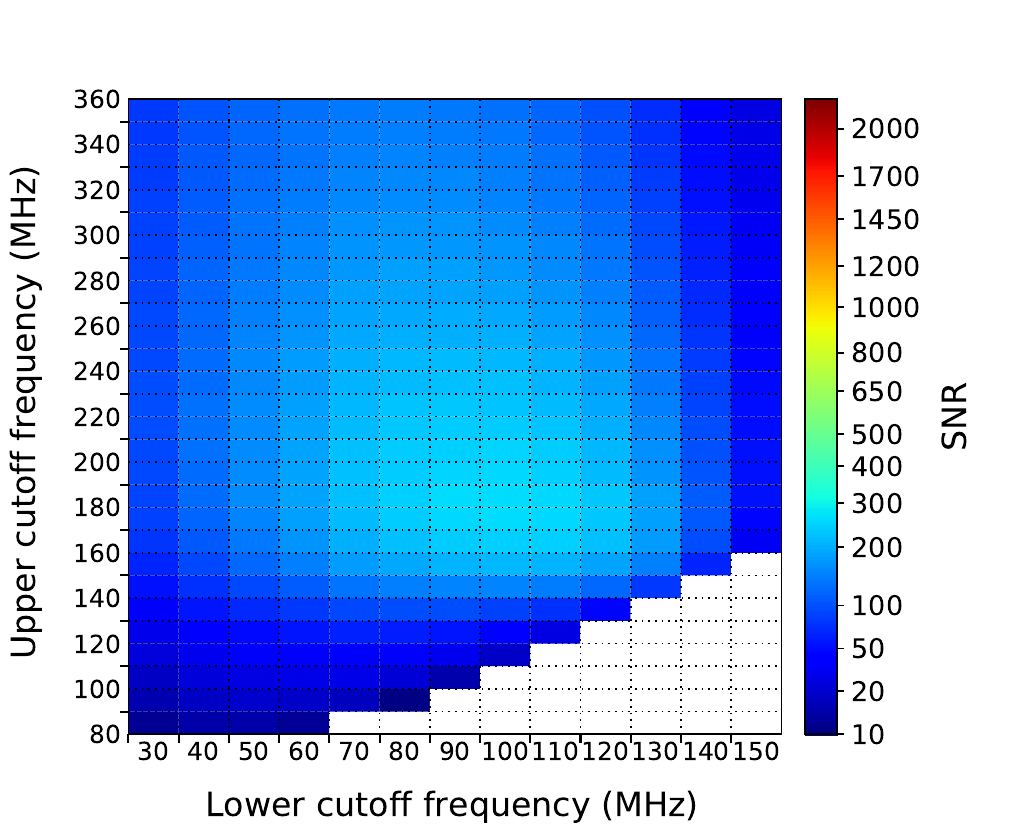}
\caption{Outside the Cherenkov ring (distance: 246 m)}

\end{subfigure}
 \setcounter{figure}{4}
\caption{SNR seen in a typical antenna inside, on and outside the Cherenkov ring respectively, at various frequency bands, for one typical shower induced by a 10 PeV gamma-ray primary 
with zenith angle $= 61^\circ$ and $\alpha = 79^\circ$. \label{fig:BWscan}}
 
\end{figure}

\begin{figure}
 \hspace{-0.5cm}
\includegraphics[width=1.05\columnwidth]{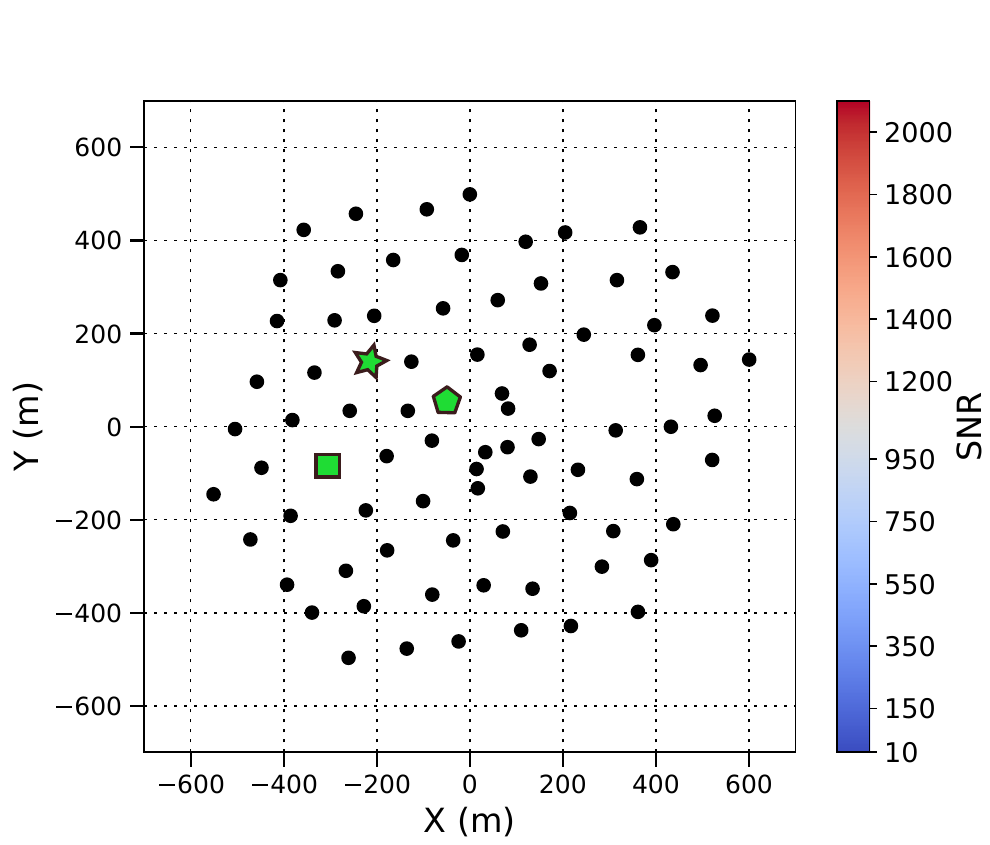}
\caption{SNR map of a 10 PeV gamma-ray shower ($\theta = 61^{\circ}$) at 100-190 MHz. The black dots represent the 81 antenna positions. The antenna on the Cherenkov ring used for the frequency band scan is shown by
the star shape. The square shape represents the antenna outside the Cherenkov ring and the pentagon that inside the Cherenkov ring.}
\label{fig:SNRmap}
\end{figure}

%
%

A scan of the possible frequency bands that can be used for the measurement of air showers of energy 10 PeV  is made. That is, we can construct a heat map of the SNR in different frequency bands. 
The frequencies for the heat map range from 30 MHz to 150 MHz for the lower edge of the frequency band and from 80 MHz to 350 MHz for the upper 
edge of the band. Such a scan is made for antenna stations at each region mentioned above. This is shown in Figure~\ref{fig:BWscan} for a typical gamma-ray shower 
with a zenith angle of $61^\circ$ and an energy of 10 PeV. 

It is obvious that the typical frequency band of 30-80 MHz (lower left bins in Figure~\ref{fig:BWscan}) is not 
ideal for obtaining an optimal level of SNR. In the figure, the brightest zone for each region on the shower footprint shows the ideal frequency band, where a maximum range of SNR is obtained.
Taking measurements at frequencies like 100-190 MHz gives a higher SNR. All bands where a value of SNR less than 10 is obtained are set to the color white, 
since this is the typical threshold for detection in an individual antenna station \cite{Aab:2015vta}.
The bands with high SNR become especially crucial, when the energy threshold is attempted to be lowered. 
A map of the SNR that is measurable by the antennas is shown in Figure~\ref{fig:SNRmap}. The black dots represent the 81 antennas considered in the simulations. The antennas considered for the frequency band scan in 
Figure~\ref{fig:BWscan} are also marked here. The SNR map shown in the figure is obtained for the frequency band 100-190 MHz\footnote{This was produced by running a CoREAS simulation 
in parallel mode with 3750 antennas using the hadronic interaction model
UrQMD instead of FLUKA}.

Showers of other zenith angles and other primaries also show a similar
behavior in the frequency band scan. As the zenith angle and the primary type changes, there is a variation 
in the scaling of SNR. This results from the change in the total electromagnetic content (for different primary type) and the different spread of the signal strength on the ground (for different zenith angle). 
There is a direct relation between the spread in the diameter of the Cherenkov ring and the inclination of the shower. 
Thus, the frequency bands with a higher value of SNR are the same for showers of other primaries and other zenith angles 
as that for a shower of zenith angle $61^\circ$ (shown in Figure~\ref{fig:BWscan}).

The observed signal-to-noise ratio in the antennas will depend on the energy of the shower, the zenith angle, and the azimuth angle (resulting in varying values of the Geomagnetic angle). The study of SNR  in these
parameter spaces is described in the following sections. The variation of the SNR with respect to the changing position of the shower maximum is not taken into account over here. 
\begin{figure}
 \captionsetup[subfigure]{justification=centering}
  \begin{subfigure}{.99\columnwidth}
 \centering

  \includegraphics[width=1\linewidth]{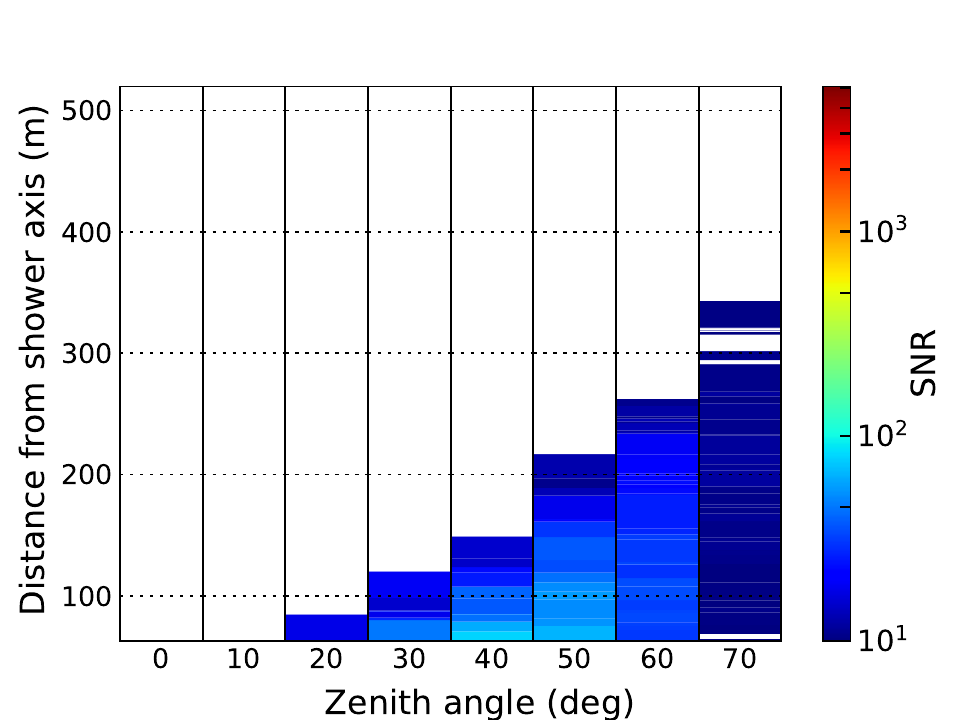}
  \caption{30-80 MHz}
 \end{subfigure}
 \begin{subfigure}{.99\columnwidth}
  \centering
   \includegraphics[width=1\linewidth]{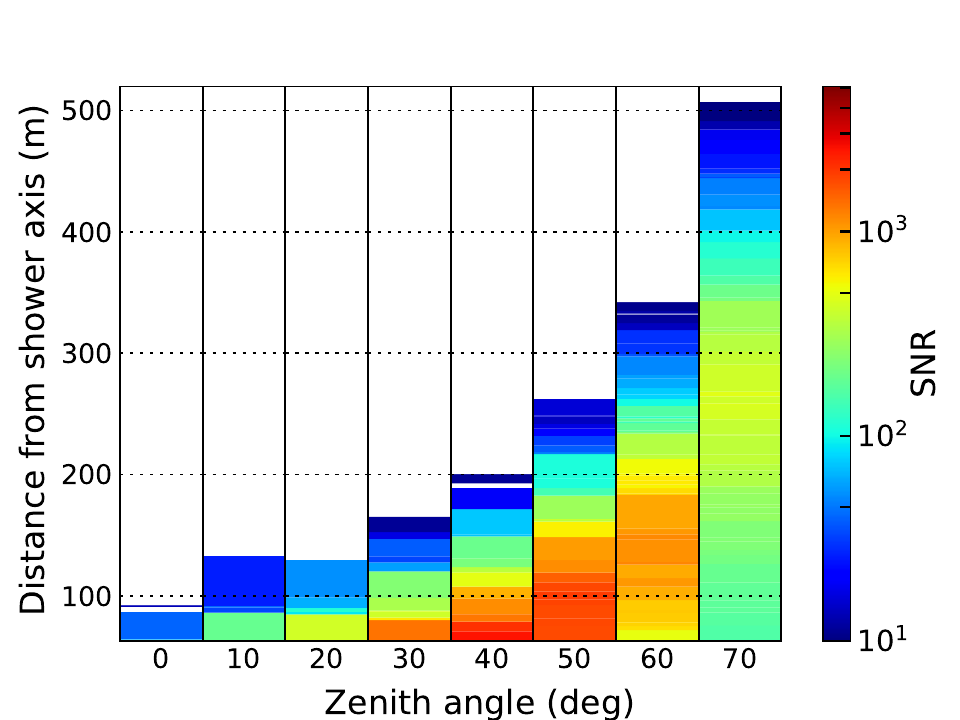}
 \caption{50-350 MHz}
  \end{subfigure}
  \begin{subfigure}{.99\columnwidth}
  \centering
   \includegraphics[width=1\linewidth]{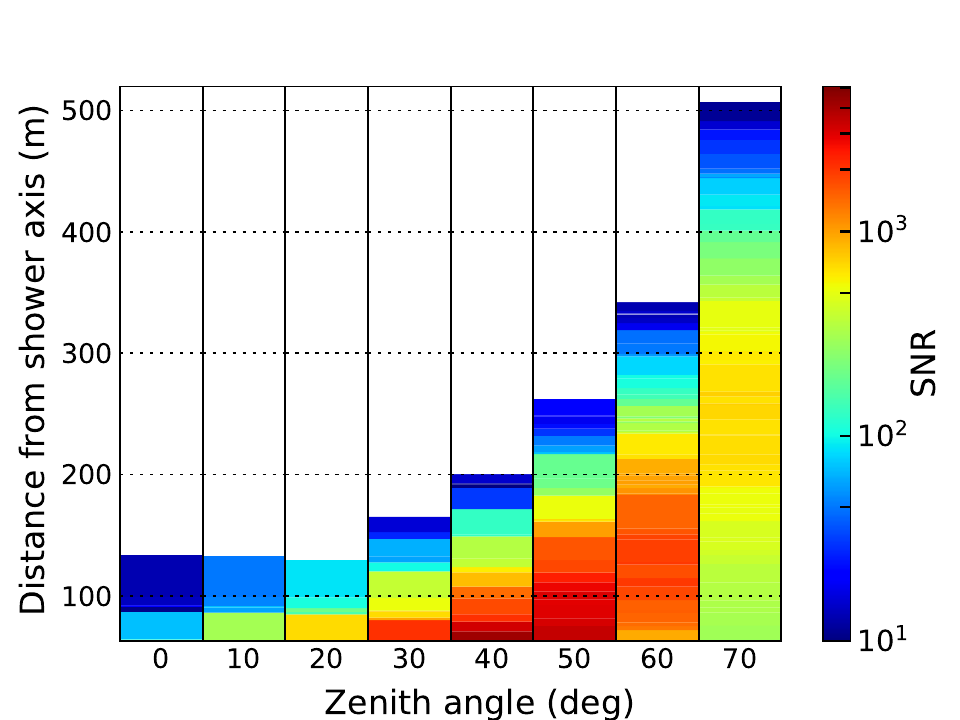}
 \caption{100-190 MHz}
  \end{subfigure}
  \caption{Zenith angle dependence of the SNR for showers produced by 10 PeV primary gamma ray with $\phi$ = 0. Each bin contains a typical shower for the respective zenith angle.
  At $\theta$ = $70^\circ$ the shower illuminates almost the entire array (b,c). \label{fig:zenith}}
 \end{figure}
 
  \begin{figure*}[tbph]
 \includegraphics[width=1\linewidth]{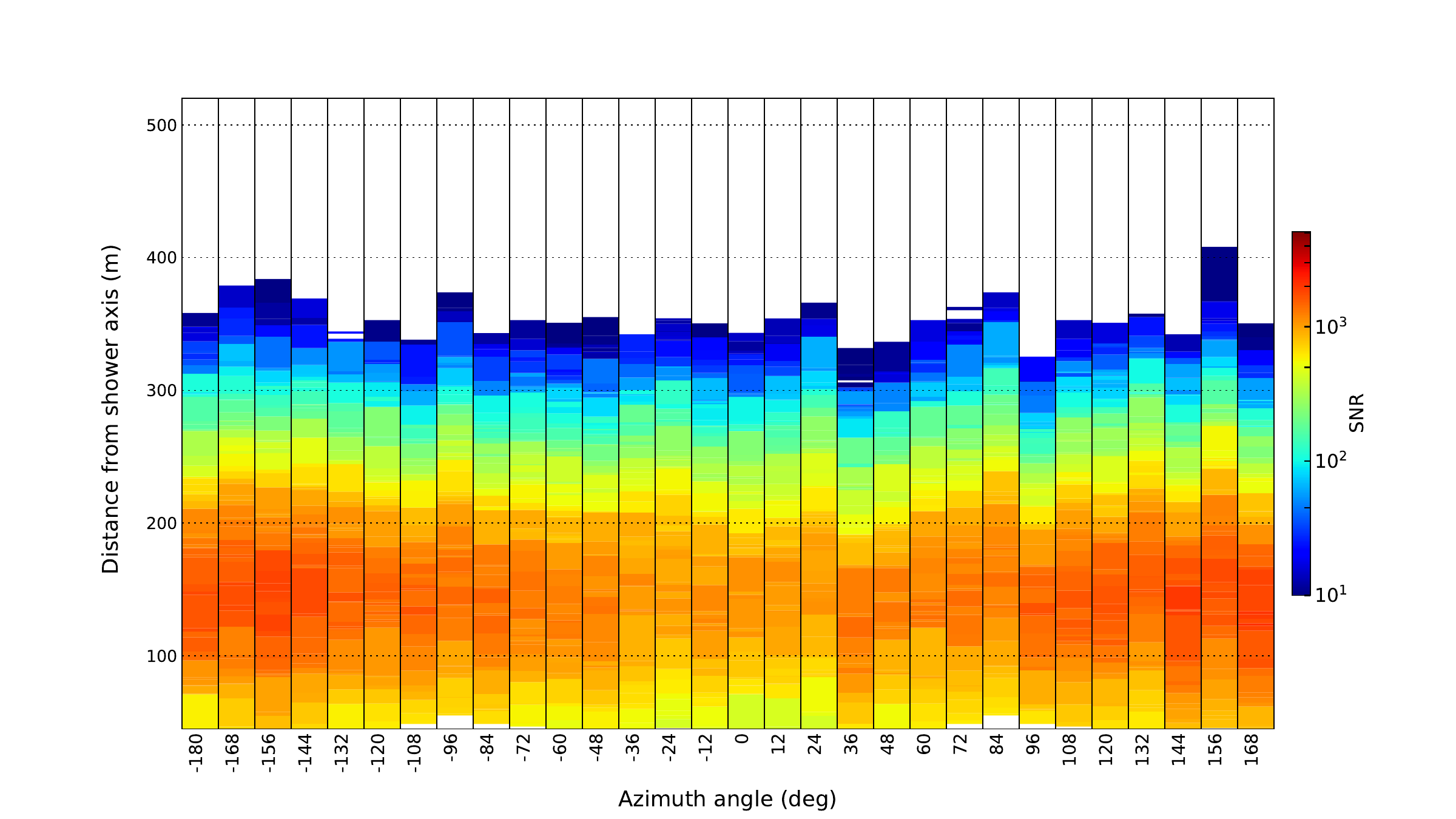}
 \caption{Azimuth angle dependence of the SNR for 10 PeV gamma-ray showers with zenith angle fixed to $61^\circ$. The figure shows a typical shower in each bin. The variation in the distance arises due to shower-to-shower 
 fluctuations and azimuth angle variations.}
  \label{fig:azimuth}

\end{figure*}
\subsection{Dependence on the zenith angle} 
\label{sec:zenith}
The evolution of the SNR with the zenith angle can be looked at for different frequency bands.
This evolution is looked at for antenna stations at various perpendicular distances to the shower axis 
(which is equivalent to the radial distance of the antennas
to the shower axis in the shower plane). Such an evolution is shown for zenith angles ranging from $0^\circ$ to $70^\circ$ in Figure~\ref{fig:zenith}, for the bands 30-80 MHz, 100-190 MHz and 50-350 MHz.

For the standard band of 30-80 MHz, the signal-to-noise ratio is significantly lower than that for the bands 50-350 MHz and 100-190 MHz.
Among all the three bands, the highest level of signal-to-noise ratio is obtained for 100-190 MHz for all zenith angles, as expected.
In particular, for showers of greater inclination, a higher signal-to-noise ratio is achieved in most of the antennas if we use the higher
frequency bands.
The areas where the Cherenkov ring falls on the antennas are visible for the higher frequencies. These are the really bright regions seen for each zenith angle and appears only for the more inclined showers.

At lower zenith angles, a major part of the shower is lost because of clipping effects. The high observation level at the South Pole is the
reason for the showers getting clipped off. The distance to the shower maximum at these zenith angles is about a few kilometers, while that
for showers of $70^\circ$ inclination is in the order of tens of kilometers. The clipping of the shower at lower zenith angles causes the
radio emission to be underdeveloped for detection. This is also the reason for the appearance of the Cherenkov ring only for zenith angles $\gtrsim 30^\circ$.

In Figure~\ref{fig:zenith}, the distances of the antennas from the shower axis fall within the range of 50 m to approximately 520 m, but only the antennas with a SNR $>10$ can detect these showers.
For vertical showers, these are the antennas with distances of $\approx$ 100 meters and for inclined showers, these are the antennas  that are even as far away as 500 m. This range 
corresponds to the required minimum spacing to detect these showers. That is, for vertical showers the antennas could at most have a spacing of 100 m and for inclined showers with $\theta \gtrsim 60^\circ$ a spacing of 300 m is 
sufficient to achieve a threshold of 10 PeV.

It is a known feature that the farther the shower maximum is from the observation level, the greater is the radius of the Cherenkov ring. 
This is purely due to geometric effects of shower propagation. The propagation of the Cherenkov ring signature in the figure
as the zenith angle increases is a manifestation of this. For an observation level of 2838 m above sea level, the average distance at which the Cherenkov ring falls is $\mathrm{d_{Ch}} \approx 250$ m for a shower of zenith angle 
$70^\circ$
and is $\mathrm{d_{Ch}} \approx 150$ m for a shower of zenith angle $60^\circ$.

The total energy fluence of the radio signal at the ground increases up to the zenith angle where clipping effects are no longer observed. 
On an average, it was seen that for 10 PeV gamma-ray showers, the total radiated energy does not get clipped-off for zenith angles greater than $50^\circ$.
For zenith angles greater than this, the total energy in the radio footprint remains 
nearly the same,
but the area increases. This results in a lower power per unit area on the ground, causing a decrease in the SNR. 
The relatively lower signal-to-noise ratio 
for the $70^{\circ}$ shower in Figure~\ref{fig:zenith} as compared to the $60^{\circ}$ shower is an effect of this.

%
 \begin{figure*}
\begin{subfigure}{.45\textwidth}
\includegraphics[width=1\linewidth]{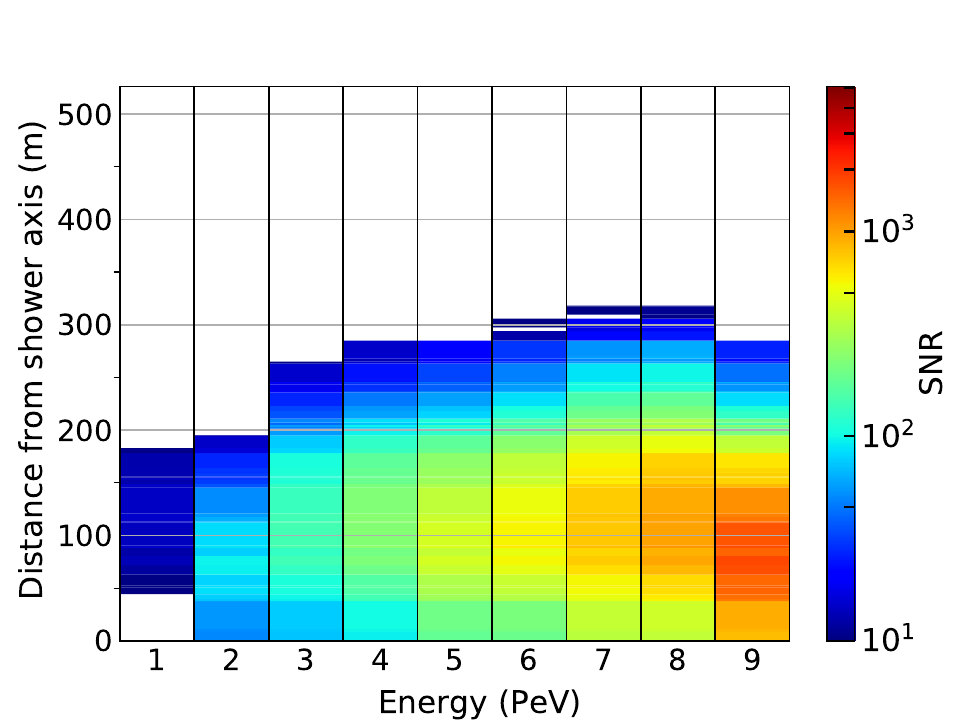}
    \caption{$\gamma$-ray, $\theta = 61^{\circ}, \phi = 0^{\circ} (\alpha =  79^{\circ})$}
    \includegraphics[width=1\linewidth]{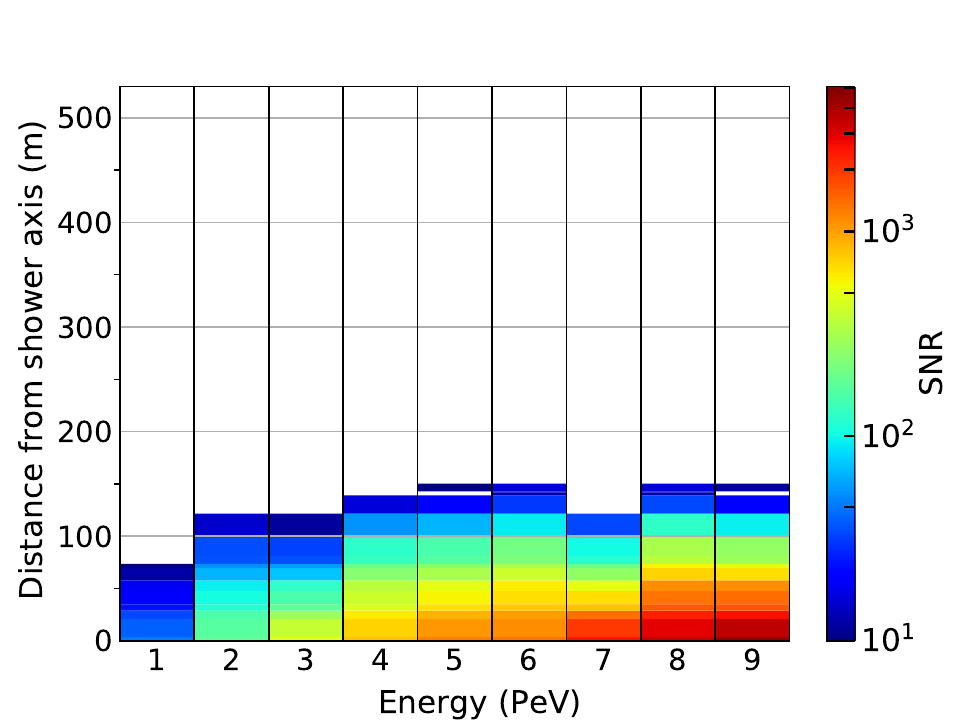}
    \caption{$\gamma$-ray, $\theta = 40^{\circ}, \phi = 0^{\circ} (\alpha =  58^{\circ})$}
     \includegraphics[width=1\linewidth]{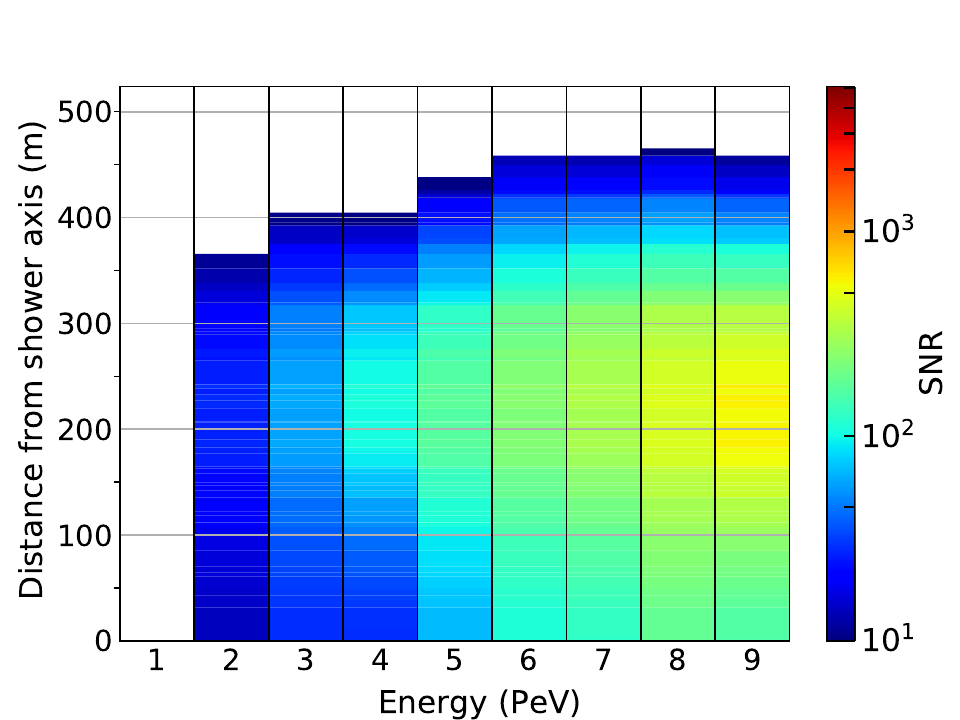}
     \caption{$\gamma$-ray, $\theta = 70^{\circ}, \phi = 0^{\circ} (\alpha =  88^{\circ})$}
\end{subfigure}
\begin{subfigure}{.45\textwidth}
\includegraphics[width=1\linewidth]{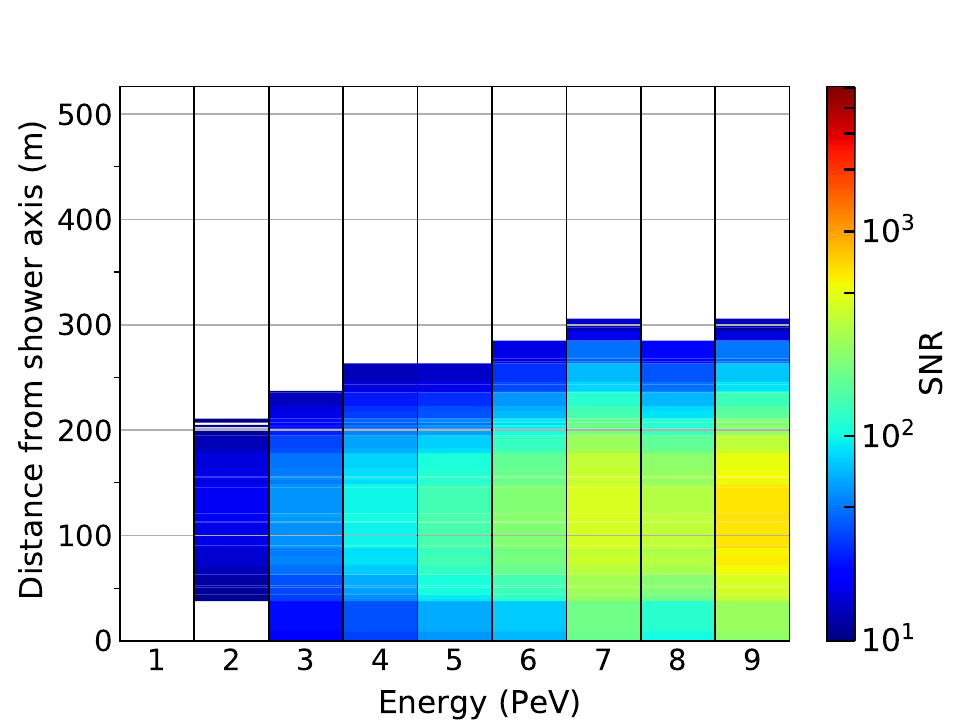}
    \caption{proton, $\theta = 61^{\circ}, \phi = 0^{\circ}( \alpha =  79^{\circ})$}
    \includegraphics[width=1\linewidth]{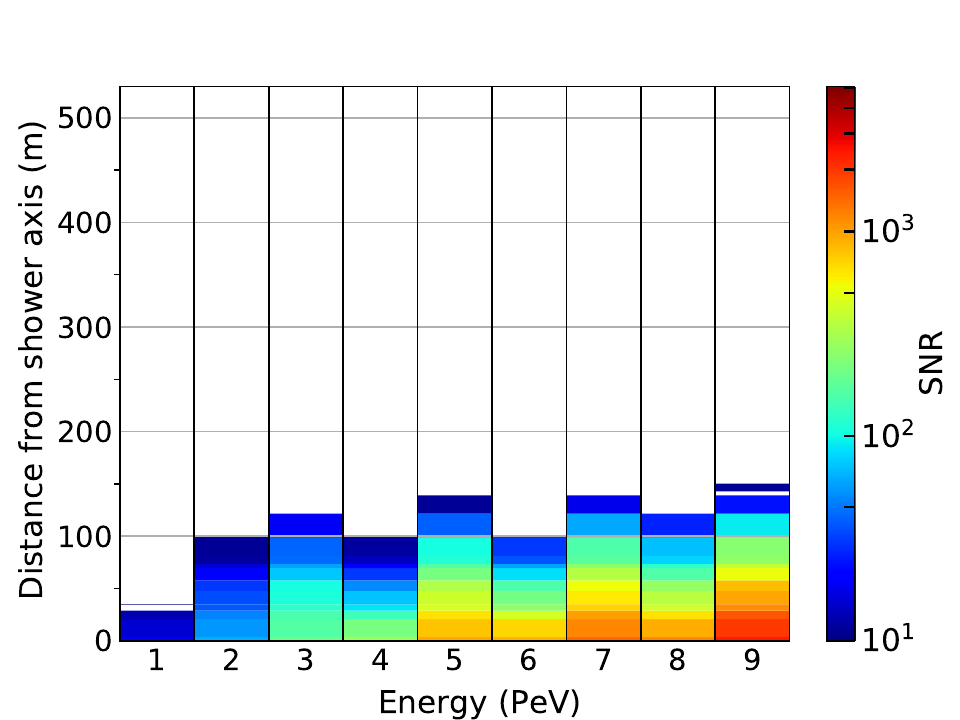}
    \caption{proton, $\theta = 40^{\circ}, \phi = 0^{\circ} (\alpha =  58^{\circ})$}
     \includegraphics[width=1\linewidth]{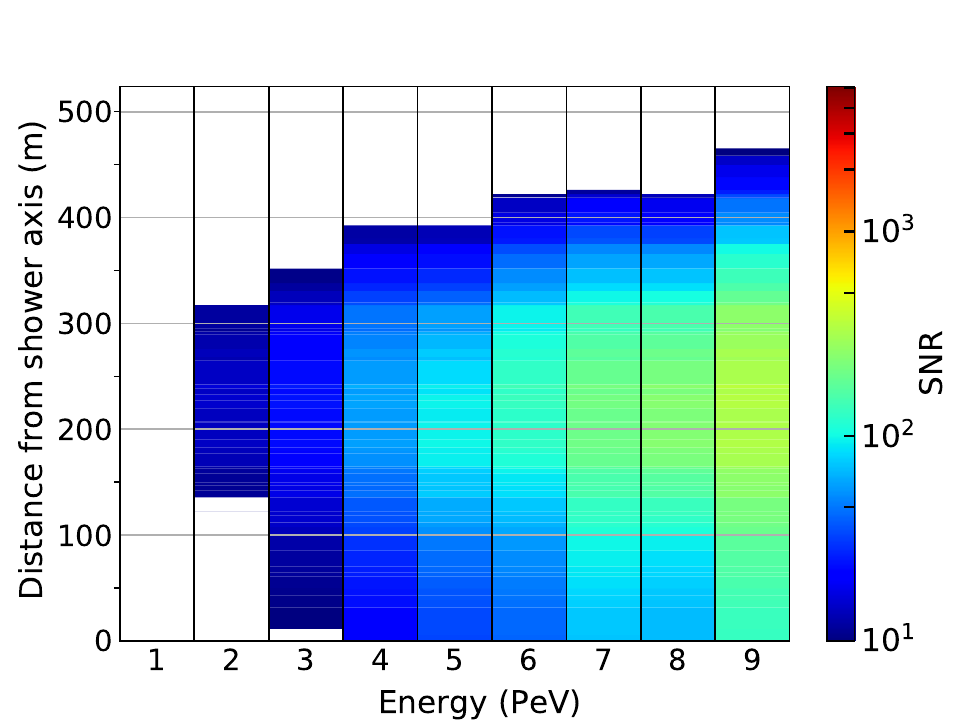}
     \caption{proton, $\theta = 70^{\circ}, \phi = 0^{\circ} (\alpha =  88^{\circ})$}
\end{subfigure}
\caption{Left: Gamma-ray induced showers at 100-190 MHz. Right: Proton induced showers at 100-190 MHz. These are typical showers at these energies. Each bin contains one sample shower for the respective energy. 
The variation in the distances where antennas with 
SNR $> 10$ are obtained for the same zenith angle arises due to shower-to-shower fluctuations. \label{fig:energy}}
\end{figure*}

%
%

\subsection{Dependence on the azimuth angle}
\label{sec:azimuth}
Another parameter that the signal-to-noise ratio depends on is the azimuth angle of the shower. Variations in the azimuth angle result in changes in the
geomagnetic angle. As a shower of zenith angle $61^{\circ}$ covers a range of azimuth angles from -180 to 180 degrees, the geomagnetic angle (at the South
Pole, where the magnetic field is inclined to the vertical direction by $18^{\circ}$) varies from $43^{\circ}$ to $79^{\circ}$. 
This leads to an amplitude  variation by a factor of $\frac{\mathrm{sin(43^{\circ})}}{\mathrm{sin(79^{\circ})}} = 0.7$.
We find that for gamma-ray showers with these range of orientations and with an energy of 10 PeV, the maximum value of the SNR varied with a standard deviation of $\mathrm{\sigma_{SNR}} = 264$ with a mean value 1518.
That is, with changing the azimuth angle there is a variation in the maximum value of the SNR by 17.4$\%$ about the mean. 
Apart from this, there is also a variation of the amplitude at a fixed azimuth angle due to shower-to-shower fluctuations which comes to 3.7$\%$ on an average. This will be further discussed in section \ref{sec:energy}.

We can thus infer that for inclined air showers at the South Pole, there is not a strong variation of the signal-to-noise ratio as the azimuth angle varies.
This is shown in Figure~\ref{fig:azimuth}. Here, gamma-ray showers each with an energy of 10 PeV
and an inclination of $61^{\circ}$ and with varying azimuth angles are shown. Thus it is justified to study the other effects only at one particular azimuth angle.

\subsection{Dependence on the primary energy} 
\label{sec:energy}
\begin{figure}[tbp]
  
 \includegraphics[width=1\linewidth]{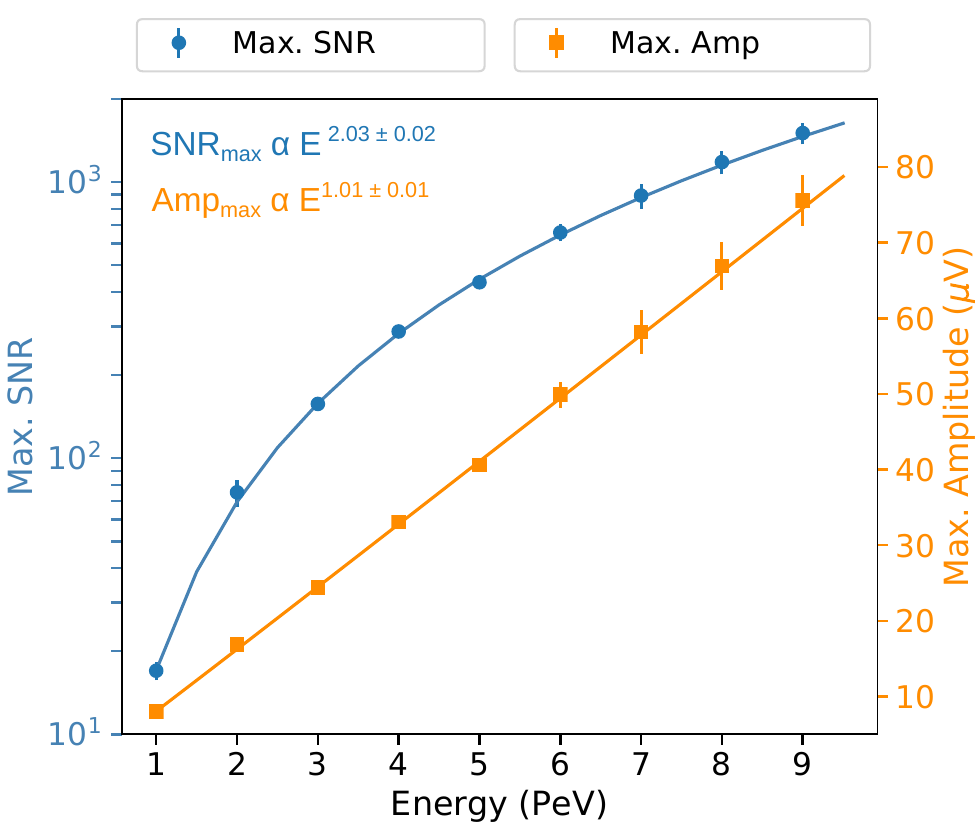}
 
 \caption{The evolution of maximum SNR and maximum amplitude for gamma-ray showers with $\alpha$ = $79^{\circ}$ and $\theta$ = $61^{\circ}$ in the frequency band 100-190 MHz. The points represent the mean value 
 and the standard deviation arising due to shower-to-shower fluctuations. The best fit to both set of simulated data points are shown, which has an $E^2$ nature for the maximum SNR and $E^1$ nature for 
 the maximum amplitude. }
  \label{fig:evolution}

\end{figure}
The signal that is observed by the antennas will obviously depend on the energy of the primary particle. The SNR becomes weaker as the 
energy of the primary particle decreases. The signal-to-noise ratio of showers with gamma-ray and proton primaries with energies ranging
from 1 PeV to 9 PeV, are shown in Figure~\ref{fig:energy}. These are showers with zenith angles of $61^{\circ}$, $40^{\circ}$ and $70^{\circ}$, and are filtered to the band 100-190 MHz. 

If we use the optimal frequency band like 100-190 MHz, we will be able to lower the threshold of detection down to 1 PeV for gamma-ray showers which have a zenith angle of $61^{\circ}$. For detection, it is required 
that a minimum of three antennas have a SNR above 10. For $61^{\circ}$ showers, we can achieve this, provided we have at least three 
antennas within a distance of  $\sim 50-180$ m from the shower axis. This is mainly the area where the Cherenkov ring falls on
the antenna array that gives a higher level of SNR. For proton showers of $61^{\circ}$ inclination, it is possible to lower the energy threshold to the level of 2 PeV in the band 100-190 MHz.

In a similar manner, the energy threshold can be lowered for showers with zenith angles $40^{\circ}$ and $70^{\circ}$ as shown in Figure~\ref{fig:energy}.
For showers with $\theta = 40^{\circ}$, we need at least three antennas within a distance of $\sim 80$ m from the shower axis. This means that a much denser
array is needed in this case. In the case of the $70^{\circ}$ showers, the minimum energy that can be detected is 2 PeV and is nearly independent of how dense the array spacing is. 

The showers shown in Figure~\ref{fig:energy} are sample showers in these energy ranges. They will also have shower-to-shower fluctuations, because of which the amplitude 
detected in each antenna station will differ. Taking such fluctuations into account, gamma-ray induced air showers with zenith angles of $61^{\circ}$ and azimuth angles of
$0^{\circ}$ were simulated with 11 simulations at each energy. Figure~\ref{fig:evolution} shows the fluctuations in the maximum SNR and the maximum amplitude
for these gamma-ray showers with energies ranging from 1-9 PeV. This is shown in the figure for a frequency range of 100-190 MHz.
These showers
were seen to have an average relative standard deviation of the maximum SNR of 7.6$\%$ for all energies. Similarly, a 3.7$\%$ variation in the maximum amplitude is obtained.

It is seen that there is a clear correlation between the maximum SNR (or maximum amplitude) obtained and the energy of the primary particle. 
The maximum SNR was seen to be proportional to $E^2$ and the maximum amplitude $\propto E$. A fit of $\mathrm{SNR_{max}}$ = (17.04 $\pm 0.43$) $\times$ $E^{2.03 \pm 0.02}$) was obtained.
Similarly, the maximum amplitude was seen to be related to the energy as $\mathrm{Amp_{max}}$ = (8.04 $\pm 0.10$) $\times$ $E^{1.01 \pm 0.01}$)

Detection of air showers using the radio technique in the PeV energy range is something that has not been achieved so far. This study shows that such a detection 
is possible if the measurement is taken in the optimum frequency range, e.g. 100-190 MHz. This means that by using this frequency range, for radio air shower detectors, 
it is possible to search for gamma rays of PeV energy arriving at the South Pole, from the Galactic Center.

Such a method can indeed be used at other locations on the Earth, for increasing the probability of detection of lower energy air shower events. 
However, the Galactic Center may not be visible at all times.
The exact threshold for 
detection may vary depending on the observation level, the magnetic field at these locations and the dimensions of the antenna array. The noise conditions of these areas will also affect the measurement. 
The use of the optimum frequencies will nevertheless increase the detection rate of inclined air showers and will lower the energy threshold.
In addition, by using interferometric methods, the very conservative condition of SNR $>$ 10 in 3 antennas can certainly be achieved.

\begin{figure}[tbp]
  \includegraphics[width=1\linewidth]{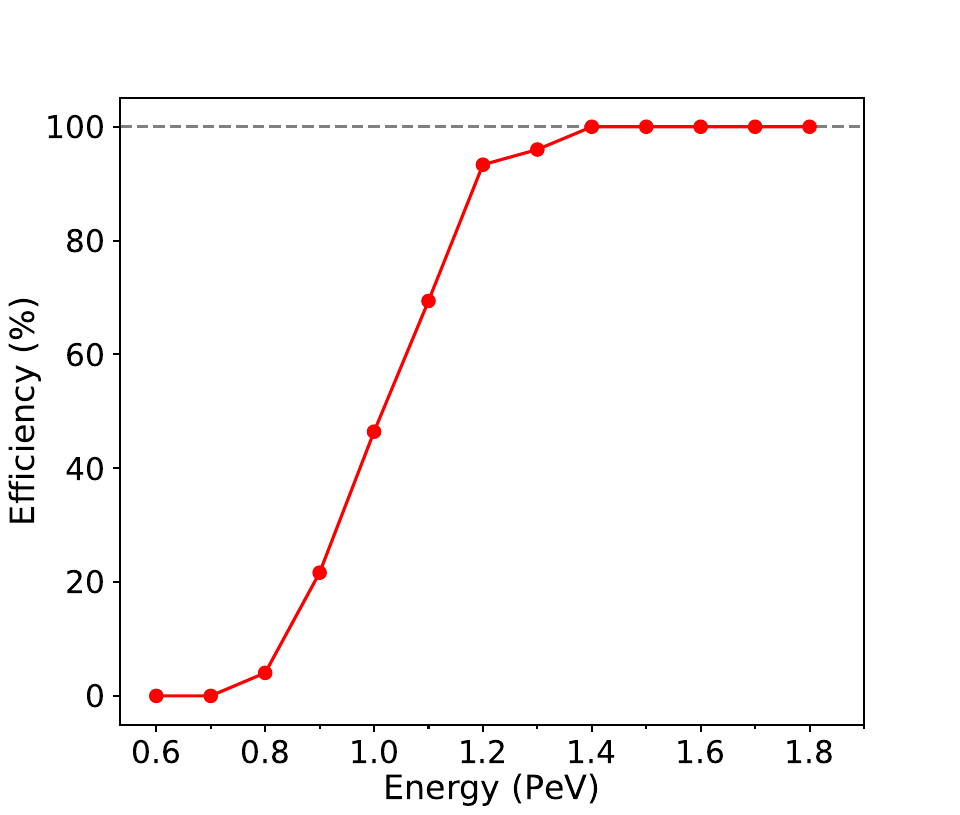}
  \caption{Efficiency of detection at different energies. An SNR $>$ 10 in at least 3 antennas is applied as the condition for detection. This is
  tested for 100 $\gamma$-ray showers with $\theta = 61^\circ$ in each energy bin in the frequency band 100-190 MHz. \label{fig:efficiency}}
 \end{figure}

\section{Efficiency of detection}

The results quoted in section~\ref{sec:energy} will depend on fluctuations between different showers of PeV energy, that arrive at the antenna array with different azimuth angles and different core positions. 
These factors, along with shower-to-shower fluctuations, will affect the rate of detection of air showers produced by PeV gamma rays. 

\begin{sloppypar}
To have an estimate of this, 170 simulations of gamma-ray induced showers with an energy of 1 PeV and zenith angle of $61^{\circ}$
were performed. These simulations had random azimuth angles and random core positions. 
Out of these showers, those with their core positions lying within a radius $\approx$ 564 m (corresponding to an area of 1 km$^2$) from the center of the array were chosen. This reduced the sample size to 140 events.
If more than three antennas in the array have SNR $> 10$, the shower is detected. 
Upon conducting this test it was seen that these showers were detected with an efficiency of 47$\%$.
\end{sloppypar}


To have a better estimate of the energy where an efficiency of 100$\%$  is reached and the energy where the efficiency goes down to 0$\%$, simulations were done from 0.6 PeV to 1.8 PeV with 100 simulations
in each energy bin, and with a bin width of 0.1 PeV. It was seen that an efficiency of 100$\%$ is reached for an energy of 1.4 PeV and the efficiency goes down
to 0$\%$ below 0.7 PeV. The efficiency curve for the simulated showers is shown in Figure~\ref{fig:efficiency}.

\section{Discussion}
As discussed in section~\ref{sec:energy} it is possible to lower the energy threshold for radio detection by using an optimum frequency band. In the case of gamma-ray showers with a zenith angle of $61^{\circ}$,
it is possible to lower the threshold down to 1 PeV for the frequency band of 100-190 MHz, if we have an antenna array with an average spacing of 125 m at the South Pole.
At other experimental locations, the threshold may vary depending on the 
specific environment of the region. This method can be used not only for the specific purpose of PeVatron detection,
but also for improving our current understanding of air showers, e.g. the study of mass composition at energies starting from the PeV range. 

The results presented here may vary depending on the exact noise that is present at the site of the experimental setup. On comparing with other available sky maps, the noise model by Cane predicts a level of 
noise that 
is slightly lower. For example, at a frequency of 110 MHz, the Cane model is seen to show around 15 $\%$ less amplitude in the level of noise than that of 
other noise models like LFmap. This will introduce
second order fluctuations in the SNR and has been neglected here. A more detailed study should also take these fluctuations into account. There will also be fluctuations depending on the local sidereal time.
A thermal noise level of 300 K is considered in this study. Today, antennas with much lower system noise are available; e.g. the SKA-LOW prototype antenna, SKALA, has a system noise of about 40 K only \cite{2015ExA....39..567D}.
The uncertainty arising from CoREAS can be estimated from the experimental tests made on CoREAS so far. 
Different air shower experiments determined CoREAS to be accurate on
an absolute scale to better than 20$\%$ at frequencies up to 80 MHz \cite{LOPES:2015eya}\cite{Apel:2016gws}.
This means that the
uncertainty in the threshold due to the use of CoREAS is 
likely smaller than 20$\%$.

The detection potential of such an antenna array will also depend on the triggering capability. Triggers provided to the antennas by the IceTop array will not be fully efficient for PeV gamma-ray showers that are 
inclined with a zenith angle of $61^{\circ}$. Triggering is possible only if a particle from the air shower hits one of the IceTop tanks. The triggering capabilities of the future scintillator array is 
not yet studied in detail. Alternatively, if self-triggering of the antenna array is used, the energy threshold will rise depending on the broad-band radio interferences at the experimental site. 
The ARIANNA experiment has demonstrated that the conditions at Antarctica can be excellent for self-triggering \cite{Barwick:2016mxm}. 


A major challenge for the detection of these gamma rays is the background cosmic-ray flux which will be much larger than the gamma-ray flux. At energies above 0.8 PeV,
a maximum of 8 gamma-ray events can be expected from the Galactic Center in one year, for a radio array with an area of 1 km$^2$. Out of these, 5 events
will be above 1.4 PeV, where we have a full efficiency of detection, assuming that events considered to be detectable will also be triggered. In order to distinguish these gamma-ray events, in a point-source scenario, 
from the background cosmic-ray events an angular resolution of 0.1$^\circ$ or better, and
a minimum gamma-hadron separation factor of 10 is required for a detection within a 5$\sigma$ confidence level in 3 years.



\section{Conclusions}
We have performed a simulation study for the detection of air showers produced by primary gamma rays of PeV energy. The focus is on showers of zenith angle $61^{\circ}$, since this is the direction from which 
PeV gamma rays
will approach the IceCube Observatory from the Galactic Center. In order to find the best measurement parameters, CoREAS simulations have been done, assuming an antenna array at the positions of the IceTop stations.

The signal-to-noise ratio received at these antennas has mainly been focused on in this analysis. A scan of the possible frequency bands within which the experiment can operate shows that there is a 
range of frequencies
within which the SNR is at the optimum level. One of these frequency bands, namely 100-190 MHz, has been used here for studying other shower dependencies. This is the first study that shows that moving 
to this 
frequency range will help in the detection of inclined air showers. It will even help in lowering the energy threshold for gamma-ray showers with a zenith angle of $61^{\circ}$ down to $\approx$ 1 PeV at the South Pole.

For a hybrid array of 1 km$^2$ area and an average antenna spacing of 125 m, with an operating frequency band of 100-190 MHz, $61^{\circ}$ gamma-ray showers of 1 PeV can be detected with an efficiency of 47$\%$. 
This number was determined for showers whose cores 
fall within 
a circular region around the array center covering an area of 1 km$^2$. 
An antenna array with an average spacing of 125 m has been used in this case.
We can reach a full efficiency above 1.4 PeV, and have a non-zero rate of detection above 0.8 PeV.
Due to the simplifications in the simulation studies, the experimentally achievable thresholds may vary by a few tens of percents. They could even be lowered further
with the usage of sophisticated hardware or by using interferometric detection techniques.
Using even the simple radio setup considered in this paper at the IceCube location will give us a chance for detecting PeV gamma rays from the Galactic Center.

\begin{acknowledgements}
We would like to acknowledge the support of the IceCube collaboration and the members of the IceCube-Gen2 group at KIT and DESY-Zeuthen. 
We would also like to thank Anna Nelles and Larissa Paul for their very useful inputs.
\end{acknowledgements}

\renewcommand\thefigure{\thesection.\arabic{figure}} 
\renewcommand\thetable{\thesection.\arabic{table}} 
\appendix

\section{Estimation of the number of gamma-ray events with PeV energies}
\label{sec:flux}
%
%
%
%
It is essential to have an estimate of the number of gamma-ray events of PeV energy expected to approach the detector from the direction of the Galactic Center. For this, we have to extrapolate the flux of TeV
gammas that has been observed by H.E.S.S.. A simple extrapolation to PeV energies, without any cut-off (which is preferred by the H.E.S.S. data points), 
is shown in Figure~\ref{fig:extrapolation}.
A spectrum with cut-off at energies 
like 1 PeV, 10 PeV or 100 PeV is also possible. Here we mainly consider the best-case scenario, that is a spectrum without any cut-off.
A spectrum of $\frac{dN}{dE}\propto E^{-2.32}$ is used for the extrapolation, which is the best fit to the H.E.S.S. data points.

\begin{figure}[H]
\centering
\includegraphics[width=1\linewidth]{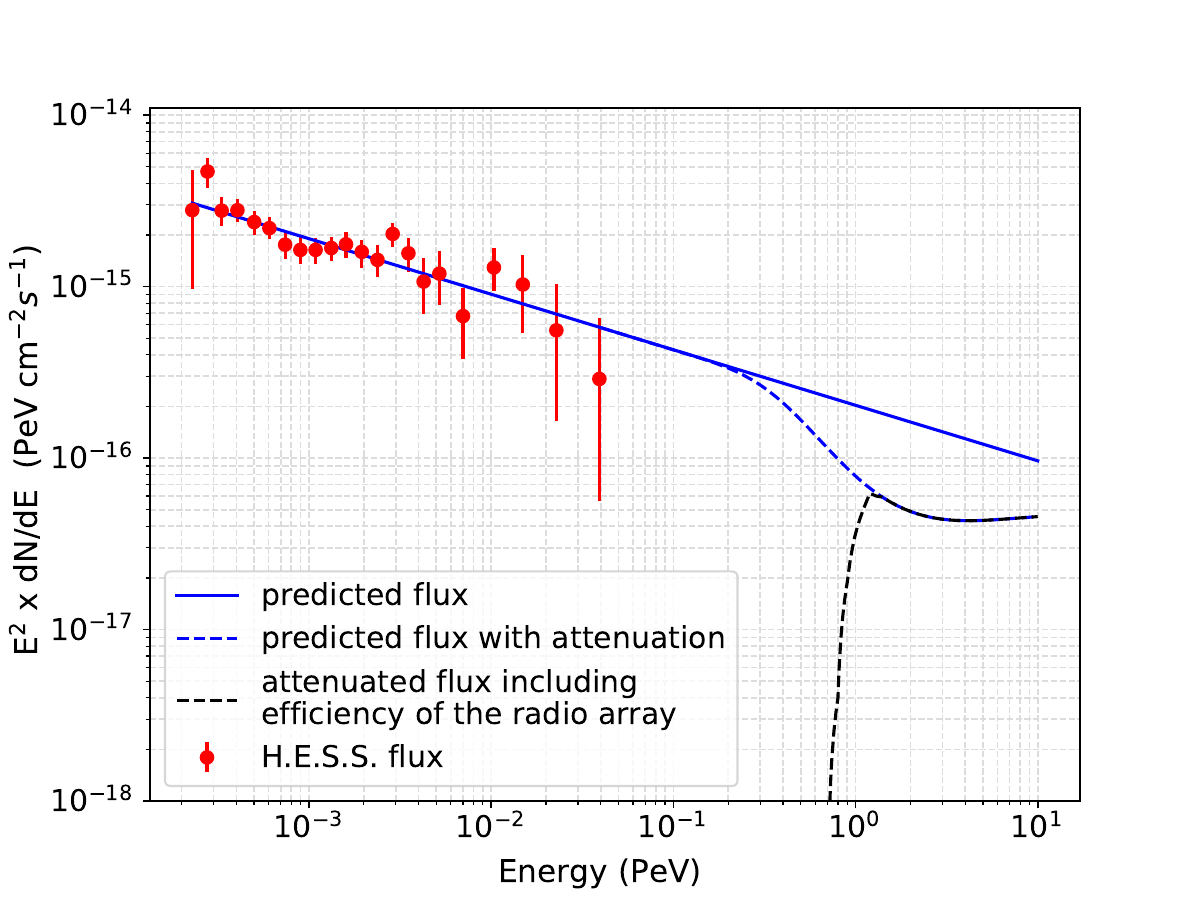}
\setcounter{figure}{0}

\caption{Flux as seen by H.E.S.S.~\cite{Abramowski:2016mir}, with a simple $E^{-2.32}$ extrapolation. The gamma-ray flux will get attenuated due to 
interactions with the CMB. The detectable flux takes the detection efficiency of the radio array into account. \label{fig:extrapolation}}.
\end{figure}

The extrapolated flux will also get attenuated due to the CMB \cite{Vernetto:2016alq}. This leads to a survival probability of the gamma rays from the extrapolated flux
given by $\approx \frac{1}{e^{L_{\mathrm{dis}}/L_{\mathrm{atten}}}}$, where $L_{\mathrm{dis}}$ is the distance traveled by the gamma rays (here it is the distance between
the Earth and the Galactic Center $\approx$ 8.5 kpc) and $L_{\mathrm{atten}}$ is the attenuation lengths of the gamma rays at different energies.
The resulting spectrum after attenuation is also shown in Figure~\ref{fig:extrapolation}.
Finally, the efficiency of detection of the radio array at various energies is also taken into consideration. The resulting flux
that will be seen by the antenna array is shown by the black curve in the figure.
From this method of extrapolation, an estimate of the expected number of events above PeV energies, detected in a year in a array of area 1 km$^2$, has been derived.
Since the Galactic Center lies at an inclination
of $61^\circ$ at the South Pole, the area of coverage of the array has to be weighted by a geometry factor of cosine($61^\circ$).


The expected number of events above 0.8 PeV, where we have a non-zero efficiency of detection, 
and above 1.4 PeV, where we have a full efficiency, are shown in Table~\ref{table:flux}. Here, we have made the assumption that the events that can be detected by the radio array will also be triggered.
The number of events before and after folding through the detection efficiency are shown in the table.
These numbers are evaluated using the attenuated flux of gamma rays. 

\begin{table}[H]
\centering
\begin{tabular}{c|c|c}
\multicolumn{1}{c|}{}  & \begin{tabular}[c]{@{}c@{}}$N\mathrm{_{events}}$($>$  $E_0$)\\   (1 yr) \end{tabular}  & \begin{tabular}[c]{@{}c@{}}$N\mathrm{_{events}}$($>$  $E_0$) \\ $\times$ efficiency \end{tabular}\\ \hline
\begin{tabular}[c]{@{}r@{}}    $E_0$ = 0.8 PeV \\ (efficiency $> 0$)\end{tabular}                 & 11.5    & 7.9   \\ \hline
\begin{tabular}[c]{@{}r@{}}    $E_0$ = 1.4 PeV \\ (full efficiency)\end{tabular}                 & 5.1     & 5.1 \end{tabular}
\caption{Estimated number of events per year obtained from the extrapolation of the attenuated gamma-ray flux, with and without detector efficiency limits. \label{table:flux}}
\end{table}

To obtain an estimate of the required gamma-hadron separation factor, we compared the expected number of gamma rays in 3 years with the number of cosmic rays 
that IceTop can see in 3 years within a region of the sky with a diameter of 0.1$^\circ$. Above 0.8 PeV, the expected number of cosmic rays is $\approx$ 289.3 and gamma rays is $\approx$ 23.8. This means that for a 
detection within a confidence level of 5$\sigma$, we will need a separation factor of $\approx$ 12.7
On the other hand, if we consider the number of events above 1.4 PeV (gamma rays $\approx$ 15.4 and cosmic rays $\approx$ 99.5), we require a separation factor
of $\approx$ 10.5. The gamma-hadron separation can be done by using the information of the shower maximum or by using the different muon content of showers from these primaries. The optimization of this requires a
separate, deeper study.

\section{Generating a noise trace}
\label{sec:noisetrace}

A model for the Galactic noise developed by Cane~\cite{Cane} is used for the following discussion. The Galactic noise is provided in units of brightness ($ B(\nu)$) and is expressed as a function of frequency. 
Assuming the source of Galactic noise to be a blackbody and hence using the Rayleigh-Jeans law, we can relate the brightness to its brightness temperature.

\begin{equation}
B(\nu) = 2k_{\mathrm{B}}T\frac{\nu^{2}}{c^{2}}  \hspace{7mm} [ \text{W}\text{m}^{-2} \text{sr}^{-1} \text{Hz}^{-1}]
\end{equation}
where $\mathrm{k_B}$ is the Boltzmann's constant and T is the brightness temperature.\\

We can add thermal noise due to the electronics of the receiving system to the brightness temperature in order to obtain the total noise temperature.
\begin{equation}
T_{ \mathrm{tot}} = T_{ \mathrm{brightness}} + T_{ \mathrm{thermal}}
\end{equation}

The electromagnetic power of the noise obtained in the frequency band $\delta \nu$ from solid angle d$\Omega$ by an antenna of effective area of $A_{ \mathrm{eff}}$ is,
\begin{equation}
 P_{\nu}(\theta,\phi) = \frac{1}{2} B(\nu) \mathrm{d}\Omega \;A_{ \mathrm{eff}}(\theta,\phi)\; \delta \nu 
\end{equation}
Here, a factor of 1/2 has to be added in order to account for the fact that the antenna can extract power only from one of the polarizations of the incoming electromagnetic wave.\\
The Poynting flux per unit frequency is obtained by integrating the brightness over the solid angle.
\begin{equation}
\begin{aligned}
S =& \int B(\nu) \mathrm{d}\Omega \hspace{7mm} [ \text{W}\text{m}^{-2} \text{Hz}^{-1}] \\
 =& \;\frac{2k_\mathrm{B}\nu^2}{c^2} \int T(\theta, \phi) \mathrm{d}\Omega
\end{aligned}
\end{equation}
The Poynting flux within the frequency interval of $\delta \nu$ is then,
\begin{equation}
\begin{aligned}
 S_{\nu} =& \;S \delta \nu \hspace{7mm} [ \text{W}\text{m}^{-2} ]\\
 =& \;\frac{2k_\mathrm{B} \nu^2 \delta \nu}{c^2}  \int T(\theta, \phi) \mathrm{d}\Omega
\end{aligned}
\end{equation}
Again, the Poynting flux extracted at the antenna is $S_{\mathrm{rec}}=\frac{S_{\nu}}{2}$ for reasons of polarization matching.\\
We can relate the Poynting flux to the electric field delivered to the antenna as,
\begin{equation}
|\overrightarrow{S}| = \frac{1}{2nZ_0}|\overrightarrow{E}|^2
\end{equation}
where $Z_0 = 376.7303$ Ohm is the vacuum impedance.\\
Taking the refractive index of air to be 1, the amplitude of the electric field at the antenna because of the Galactic noise can then be obtained from the Poynting flux as,
\begin{equation}
\begin{aligned}
|\overrightarrow{E}|=& \sqrt{S_{\mathrm{rec}}2Z_0} \hspace{7mm} [ \text{V/m}]\\
 =& \sqrt{\frac{1}{2}2Z_0\frac{2k_\mathrm{B}\nu^2 \delta \nu}{c^2} \int T(\theta, \phi) \; \mathrm{d}\Omega}
\end{aligned}
\end{equation}
Thus, the voltage developed at the antenna is 
\begin{equation}
V(\nu) = \overrightarrow{E}\cdotp \overrightarrow{l_{\mathrm{eff}}} \hspace{7mm} [ \text{V} ]
\end{equation}

Since we have already taken into account that the polarization should match, we can multiply the modulus of the field and the modulus of the antenna height to obtain the voltage.
Of course, this simplification cannot be done for a noise model with directional dependence. 
The received voltage is now given by

 \begin{equation}
 V(\nu)  = \sqrt{2Z_0\frac{\mathrm{k_{B}} \nu^2 \delta \nu}{c^2}   \int T(\theta,\phi)\; |\overrightarrow{l_{\mathrm{eff}}}(\theta, \phi)|^2 \; \mathrm{d}\Omega}
 \end{equation}
Since the model used has the temperature to be independent of $\theta$ and $\phi$, $T(\theta,\phi)=T$ 
can be taken out of the integral.\\

The amplitude extracted from the model has no phase information of the incoming noise.
We can add random phases to the amplitude since noise indeed behaves randomly.
\begin{equation}
\begin{aligned}
  V(\nu) = V(\nu) * \mathrm{exp}(-i\varphi)
  \end{aligned}
\end{equation}
$\varphi$ is a random number that is generated between 0 and 2$\pi$.

Finally, we can convert the amplitude to the time domain using Inverse Fourier Transform: 

\begin{center}
\begin{equation}
    V(\nu)\longrightarrow V(t)
\end{equation}

\end{center}

Of course, this is only an average behavior of the noise. One can also assume variations in the extracted amplitude about this average noise. 

\bibliographystyle{unsrtnat}
\bibliography{MainDoc} 
\end{document}